	\documentclass[10pt,final,doubleside]{IEEEtran}
	
\normalsize
\usepackage{pifont,bm,multicol,amsfonts,amsmath,color,amssymb,graphicx,amsthm, epsfig,cite,psfrag,subfigure,algorithm,enumerate,stfloats,algorithm,algorithmic,epstopdf,balance}

\newtheorem{remark}{\underline{Remark}}

\begin{document}
\title{Joint Channel Estimation and Data Detection for Hybrid RIS aided Millimeter Wave OTFS Systems}

\author{
Muye Li, Shun Zhang, \emph{Senior Member, IEEE}, Yao Ge, \emph{Member, IEEE}, \\ Feifei Gao, \emph{Fellow, IEEE}, Pingzhi Fan, \emph{Fellow, IEEE}

\thanks{
Manuscript received 23 January, 2022; revised 27 May, 2022 and 27 July, 2022, accepted 6 August, 2022; date of current version 9 August, 2022.
This work of M. Li and S. Zhang was supported by the National Natural Science Foundation of China under Grant (61871455, 61901329).
The work of P. Fan was supported by NSFC Project (No.62020106001).
\emph{(Corresponding author: Shun Zhang.)}
}

\thanks{M. Li, S. Zhang are with the State Key Laboratory of Integrated Services Networks, Xidian University, Xi'an 710071, P. R. China (e-mail: myli$\_$96@stu.xidian.edu.cn; zhangshunsdu@xidian.edu.cn).}

\thanks{Y. Ge is with Continental-NTU Corporate Lab, Nanyang Technological University, SG 637553, Singapore. (e-mail: yao.ge@ntu.edu.sg).}

\thanks{F. Gao is with Department of Automation, Tsinghua University, State Key Lab of Intelligent Technologies and Systems, Tsinghua University, State Key for Information Science and Technology (TNList) Beijing 100084, P. R. China (e-mail: feifeigao@ieee.org).}

\thanks{P. Fan is with the Southwest Jiaotong University, Chengdu 611756, P. R. China (e-mail: pzfan@swjtu.edu.cn).}

}

\maketitle
\vspace{-0mm}
\begin{abstract}
For high mobility communication scenario, the recently emerged orthogonal time frequency space (OTFS) modulation introduces a new delay-Doppler domain signal space, and can provide better communication performance than traditional orthogonal frequency division multiplexing system.
This article focuses on the joint channel estimation and data detection (JCEDD) for hybrid reconfigurable intelligent surface (HRIS) aided millimeter wave (mmWave) OTFS systems.
Firstly, a new transmission structure is designed.
Within the pilot durations of the designed structure, partial HRIS elements are alternatively activated.
The time domain channel model is then exhibited.
Secondly, the received signal model for both the HRIS over time domain and the base station over delay-Doppler domain are studied.
Thirdly, by utilizing channel parameters acquired at the HRIS, an HRIS beamforming design strategy is proposed.
For the OTFS transmission, we propose a JCEDD scheme over delay-Doppler domain.
In this scheme, message passing (MP) algorithm is designed to simultaneously obtain the equivalent channel gain and the data symbols.
On the other hand, the channel parameters, i.e., the Doppler shift, the channel sparsity, and the channel variance, are updated through expectation-maximization (EM) algorithm.
By iteratively executing the MP and EM algorithm, both the channel and the unknown data symbols can be accurately acquired.
Finally, simulation results are provided to validate the effectiveness of our proposed JCEDD scheme.

\end{abstract}

\maketitle
\thispagestyle{empty}


\section{Introduction}

Millimeter wave (mmWave) has become an important technology in the fifth generation wireless communication \cite{mmwave1, mmwave2}.
Compared with traditional lower-frequency communication, its broad frequency spectrum can provide tremendous gain for communication system efficiency \cite{mmwave3, mmwave4}.
However, due to its short wavelength, electromagnetic waves in mmWave band are very sensitive to physical blockages \cite{mmwave5,AIRIS}.
Fortunately, a newly proposed technology named reconfigurable intelligent surface (RIS) can perfectly address this problem.
With the deployment of RIS, electromagnetic waves can be designed to arrive its receiver by configuring the reflection coefficient of RIS elements \cite{RIS_Mag,RIS_zhong}.
In other word, RIS can help establish a virtual line of sight (LoS) path between two terminals, and provide further improvement for system throughput.
Moreover, RIS does not require complicated signal processing units for such advantages.
Compared with relay system, the RIS aided system can achieve higher energy efficiency \cite{RIS_relay_comp}.
Hence, it is considered to have potential of wide applications in future communication networks \cite{RIS1}.


However, this enhancement from RIS is deeply dependent on channel state information (CSI).
With precise knowledge of CSI, one can further apply potential advanced signal processing techniques for better transmission performance, such as beamforming design and data detection \cite{RIS_beamforming1, RIS_beamforming2, RIS_beamforming3, RIS_data_detec}.
Since RIS is typically working in the passive mode, the CSI is usually estimated at terminals rather than at RIS.
In \cite{cascaded_CE1}, Wei \emph{et al.} proposed two iterative channel estimation techniques for RIS-empowered multi-user uplink (UL) communication system, where the parallel factor decomposition of the received signal was considered.
Liu \emph{et al.} formulated the RIS channel estimation problem as a matrix-calibration based matrix factorization task, and proposed a message-passing algorithm for the factorization of the cascaded channel in \cite{cascaded_CE3}.
In \cite{cascaded_CE4}, Hu \emph{et al.} proposed a location information aided multiple RIS system for the estimation of effective angles from the RIS to the users, and then proposed a low-complexity beamforming scheme for both the transmit beam of the BS and phase shift beam of the RIS.
Apart from the works with passive RIS, a new RIS structure called the hybrid RIS (HRIS) was recently proposed in \cite{HRIS1}.
The elements on HRIS can not only reflect the impinging signal, but also absorb it with an absorption factor \cite{HRIS2}.
When the HRIS is equipped with radio frequency (RF) chains, signal processing techniques for the absorbed signal can be executed on HRIS rather than the BS \cite{HRIS3,HRIS4}.
Thus, the elements connected with RF chains are ``activated''.
Equipped with such signal processing ability, HRIS can also configure its phase shift matrix by itself, and does not require any input control signal.
In \cite{RIS_beamforming2}, Zhang \emph{et al.} resorted to HRIS, and proposed a sparse antenna activation based channel extrapolation strategy.
By optimizing the active RIS element pattern, the full one-hop channel can be inferred from the sub-sampled channel.
The authors of \cite{RIS_beamforming2} also designed a beam searching scheme to obtain the optimal RIS phase shift matrix from the sub-sampled channel.

However, with the development of high-speed transportation and the grown communication demands, user mobility has become an important issue that cannot be ignored in communication design \cite{time_varying1, time_varying3}.
When considering mobility scenario, the above mentioned transmission schemes will face unavoidable challenges due to the limited channel coherence time and the variation of scattering  environment.
Recently, orthogonal time frequency space (OTFS) modulation is proposed for transmission in high mobility scenario \cite{OTFS1}.
Different from traditional  orthogonal frequency division multiplexing (OFDM) system, time-varying channel can be represented by a small number of static parameters over delay-Doppler domain in OTFS.
In the mean time, pilot and data symbols can also be mapped and demapped over this newly constructed two-dimensional signal space \cite{OTFS3, OTFS_Dai,ISAC_yuan}.
Compared with OFDM system, it has been illustrated that OTFS is indeed more insensitive to the magnitude of Doppler frequency shifts \cite{OFDM_OTFS_comp}.
Thus, signal processing techniques can be implemented over delay-Doppler domain for better performance with low complexity in high mobility scenario.
Liu \emph{et al.} proposed an uplink-aided downlink channel estimation scheme for massive MIMO-OTFS system in \cite{OTFS_liu}, where an expectation-maximization (EM)  based variational Bayesian framework was adopted to recover the uplink channel parameters.
In \cite{OTFS_li}, Li \emph{et al.} proposed a path division multiple access scheme for massive MIMO-OTFS networks, where the angle-delay-Doppler domain channel estimation, 3D resource space allocation and mapping, and data detection over both UL and DL were considered.

Note that \cite{OTFS_liu} and \cite{OTFS_li} relied on OFDM-based OTFS system, where a cyclic prefix (CP) is inserted before every OFDM symbol within each OTFS block, leading to a low spectrum efficiency.
For better spectral efficiency, Raviteja et al. derived an input-output relationship over delay-Doppler domain by only inserting one CP for the whole OTFS block in \cite{OTFS2}.
Based on such OTFS input-output relationship, Ge et al. proposed two efficient message passing (MP) equalization algorithms for data detection with a fractionally spaced sampling approach in \cite{OTFS_Ge1}.
In \cite{OTFS_detection2}, Shan \emph{et al.} proposed a low-complexity and low-overhead OTFS receiver by using a large-antenna array.
Notice that the data detection strategies always require accurate knowledge of CSI, which can be acquired in two ways in general.
One is to estimate channel before data transmission \cite{OTFS_liu,OTFS_li,OTFS_Ge1}, which will lead to huge occupation of channel coherent time.
The other one is to separate the pilot area and data area within one OTFS block by using a large number of guard symbols \cite{OTFS2, OTFS_detection2}, which will lower the spectral efficiency of OTFS system.
Although \cite{Superimpose} and \cite{Superimpose2_weijie_yuan} proposed superimposed pilots and data based strategy to jointly estimate the channel and recover the data symbols,
the restriction on the signal-to-interference-plus-noise ratio will also incur additional limits on its utilization.

Facing with these problems, we propose a new scheme to jointly estimate the channel and detect data symbols.
Firstly, we illustrate the configuration and a transmission scheme for the HRIS aided mmWave communication system over high mobility scenario.
In the designed transmission scheme, there are a preamble and some short pilot sequences accompanying with a number of OTFS blocks, and the HRIS is activated only within the pilot durations.
Then, the HRIS aided channel model over time domain is presented.
Besides, received signal model for both the HRIS and the BS are introduced.
By utilizing the received pilot signal, the CSI between the user and the HRIS can be easily acquired and updated at the HRIS.
Thus, we propose to design and calibrate the HRIS beamforming matrix for each OTFS blocks.
Moreover, we formulate a joint channel estimation and data detection (JCEDD) problem based on probabilistic graphical model.
For the channel gain estimation and data detection, we propose a MP based strategy;
On the other hand, the channel parameters are updated through EM algorithm.
By iteratively implementing MP and EM algorithm, the CSI of the cascaded channel and the transmitted data symbols are simultaneously obtained.

The rest of this paper is organized as follows.
Section II illustrates the system configuration and transmission scheme for the HRIS aided mmWave communication system in high mobility scenario.
Then, the received signal models for the HRIS and BS over different signal spaces are respectively described in Section III.
In addition, an HRIS beamforming design strategy is proposed.
Section IV introduces the proposed JCEDD scheme with MP and EM algorithm.
Simulation results are provided in Section V, and conclusions are drawn in Section VI.
The Appendix contains some detailed proofs at the end of the paper.

Notations: Denote lowercase (uppercase) boldface as vector (matrix).
$(\cdot )^H $, $(\cdot )^T $, $(\cdot )^{*} $, and $(\cdot )^{\dagger} $ represent the Hermitian, transpose, conjugate, and pseudo-inverse, respectively.
$\mathbf I_N $ is an $N \times N $ identity matrix.
$\mathbb E \{\cdot \} $ is the expectation operator.
Denote $|\cdot | $ as the amplitude of a complex value.
$[\mathbf A]_{i,j} $ and $\mathbf A_{\mathcal Q,:}$ (or $\mathbf A_{:, \mathcal Q} $) represent the $(i,j) $-th entry of $\mathbf A $ and the submatrix of $\mathbf A $ which contains the rows (or columns) with the index set $\mathcal Q $, respectively.
$\mathbf x_{\mathcal Q} $ is the subvector of $\mathbf x $ built by the index set $\mathcal Q $.
The real component of $x $ is expressed as $\Re \{x\}$.
$\text{diag} (\mathbf x)$ is a diagonal matrix whose diagonal elements are formed with the elements of $\mathbf x $.

\section{HRIS aided MmWave OTFS System Model}

\subsection{System Configurations and Transmission Scheme}
We consider a RIS aided mmWave MIMO system in high mobility scenario, which consists of one base station (BS), one HRIS, and one user with high mobility.
The BS is equipped with a uniform linear array (ULA) with $N_b$ antennas,
and the HRIS has $N_r$ passive elements, which are arranged as a $N_x \times N_y$ uniform planar array (UPA).
It can be checked that $N_r = N_xN_y$.
For simplicity, we assume $N_x = N_y$.
Besides, the HRIS is equipped with $N_{RF}$ RF chains.
The high mobility user with single antenna is distributed in the serving area of the HRIS.

Since the HRIS and the BS are usually placed very high, there will be limited local scatterer between them.
Hence, it can be assumed that the LoS path has dominant power within the channel between them.
However, since there are some scatterers around the user, we consider multiple paths between the user and the HRIS.
Since the BS and the HRIS are always pre-placed, we assume that their locations are perfectly known without loss of generality \cite{known_loc}.
Moreover, an analog precoder/combiner is equipped at the BS.
Generally, the direct scattering path from BS to user tends to be blocked by the possible obstacles such as buildings and trees. 
Hence, we ignore the direct link between BS and user, and focus on the HRIS assisted link.
%

\begin{figure*}[htbp]
 \centering
 \includegraphics[width=130mm]{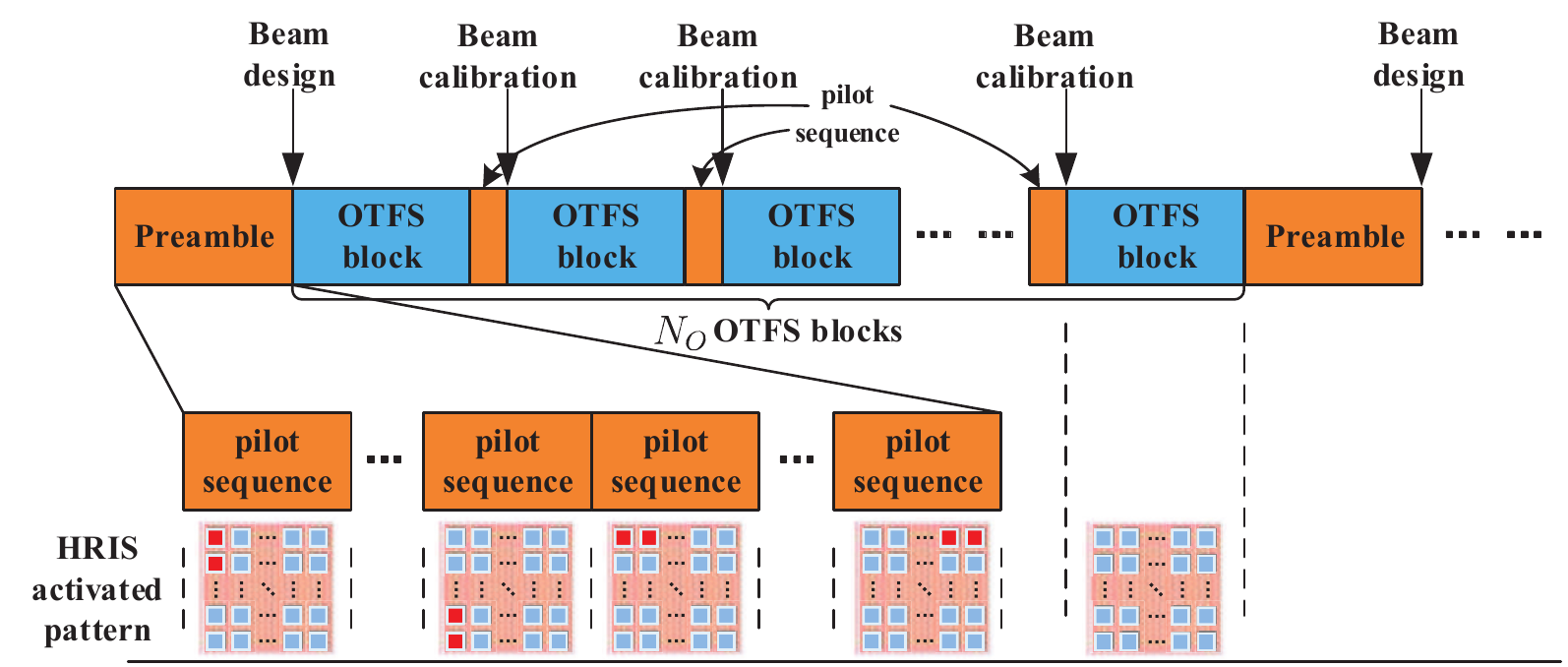}
 \caption{The illustration of the transmission structure and the HRIS activation strategy.}
 \label{transmission_structure}
\end{figure*}

In addition, to fully exploit the advantage of HRIS, we design a transmission structure, which is illustrated in \figurename{ \ref{transmission_structure}}.
The user will first send a preamble, and then execute the OTFS transmission for a relatively long time.
During the preamble, the HRIS will be in partial active mode and totally absorb the impinging signals, which means that the BS will not receive it.
Note that only the impinging signal at $N_{RF}$ activated HRIS elements can be utilized for
signal processing, i.e, channel parameter extraction.
By utilizing the acquired CSI of the user side and the known CSI of the BS side, the phase shift matrix of the HRIS, also known as HRIS beamforming matrix, can be well designed.
Besides, to assure the accuracy of parameter extraction, $2N_x$ HRIS elements should be alternatively activated at least.
Hence, the preamble is divided into $N_B = 2N_x/ N_{RF}$ time blocks.
For each time block, $N_{RF}$ HRIS elements are activated, and the user utilize the same pilot sequence $\mathbf t = [t_0, t_1,\ldots, t_{N_T-1}]^T \in \mathbb C^{N_T \times 1}$, where $N_T$ is the length of each sequence.
After the preamble, the user will adopt the OTFS modulation, where the data and pilot symbols are simultaneously transmitted to the BS.
Within the duration of OTFS blocks, the HRIS will be in full-passive mode, and will reflect the impinging signal without any absorption.
Moreover, in order to maintain the transmission quality, the user will send one pilot sequence to the HRIS between each OTFS block for the calibration of the beamforming matrix.

\subsection{Channel Model over Time Domain}
In this paper, we examine the UL transmission.
The time-frequency selective channel from user to HRIS can be expressed as
\begin{align}
\mathbf{h}^{\text{UR}}_i(t) = \sum_{p=1}^{P} h^{\text{UR}}_{p} e^{\jmath 2 \pi \nu^{\text{UR}}_{p} (t - \tau^{\text{UR}}_{p})} &\delta(iT_s-\tau^{\text{UR}}_{p}) \mathbf{a}_{\text{R}} (\phi_p,\psi_p),
\notag\\
&i = 0, \ldots, I_U - 1,
\end{align}
where $i$ denotes the index along delay domain, $I_U$ is the maximum number of delay taps between the user and the HRIS \cite{OTFS_Ge1}.
$P$ is the number of scattering paths from the user to the HRIS,
$T_s$ is the system sampling period,
$h^{\text{UR}}_{p} \sim \mathcal{CN}(0,\lambda_p^{\text{UR}})$ is the complex channel gain with its variance $\lambda_p^{\text{UR}}$,
and $\delta(\cdot)$ is a Delta function.
$\tau^{\text{UR}}_p$ and $\nu^{\text{UR}}_p$ are the delay and Doppler shift of $p$-th path, respectively.
In addition, $\phi_{p}$ and $\psi_{p}$ are the elevation and azimuth angle of arrivals (AOAs) at the HRIS.
The array response vector of HRIS
is given by
$
\mathbf{a}_{\mathbf{R}}(\phi, \psi)\!=\!\mathbf{a}_{\text{x}}(\phi, \psi) \otimes \mathbf{a}_{\text{y}}(\phi, \psi)
$,
where $\otimes$ denotes the Kronecker product.
$\mathbf{a}_{\text{x}}(\phi, \psi) \in \mathbb C^{N_x \times 1}$ and
$\mathbf{a}_{\text{y}}(\phi, \psi) \in \mathbb C^{N_y \times 1}$
are respectively the array response vectors along the elevation and the azimuth dimensions.
The $n_x$-th element of $\mathbf{a}_{\text{x}}(\phi, \psi)$ and the $n_y$-th element of $\mathbf{a}_{\text{y}}(\phi, \psi)$ are respectively defined as $[\mathbf{a}_{\text{x}}(\phi, \psi)]_{n_x} = e^{-\jmath 2\pi(n_x-1)\frac{d}{\lambda}\sin(\phi)\sin(\psi)}$ and
$[\mathbf{a}_{\text{y}}(\phi, \psi)]_{n_y} = e^{-\jmath 2\pi(N_y-1)\frac{d}{\lambda}\sin(\phi)\cos(\psi)}$,
where $d = \frac{\lambda}{2}$ is the inter-element spacing, and $\lambda$ is the signal wavelength.

As for the quasi-static flat fading channel from the HRIS to the BS, it can be written as
\begin{align}
\mathbf{H}^{\text{RB}}_{i}(t)= h^{\text{RB}} \delta(iT_s - \tau^{\text{RB}}) \mathbf{a}_{\text{B}}(\theta_{B})
&\mathbf{a}^H_{\text{R}} (\phi_{r}, \psi_{r}),
\notag \\
&i = 0, \ldots, I_B - 1,
\label{eq:H_BR}
\end{align}
$h^{\text{RB}}\sim \mathcal{CN}(0,\lambda^{\text{RB}})$ denotes the complex path gain with the variance $\lambda^{\text{RB}}$, $\tau^{\text{RB}}$ is the delay of the LoS path,
$I_B$ is the maximum number of delay taps between the HRIS and the BS, and
$\theta_{B}$ is the AOA at BS;
$\phi_{r}$ and $\psi_{r}$ are the elevation and azimuth angle of departures (AODs) at the HRIS.
Similar to $\mathbf{a}_{\text{x}}(\phi, \psi)$ and
$\mathbf{a}_{\text{y}}(\phi, \psi)$ , the array steering vector at the BS is defined as $\mathbf{a}_{\text{B}}(\theta)\in\mathbb{C}^{N_b\times 1}$ with $[\mathbf{a}_{\text{B}}(\theta)]_{n_b} = e^{-\jmath 2\pi(n_b-1)\frac{d}{\lambda}\cos(\theta)}$.

\subsection{BS Combiner Design}


Since the AOA at BS can be derived from the location information of the BS and the HRIS, an $1\times N_b$ combiner is enough for the BS.
To simplify the derivations and design, we set a Cartesian coordinate system and fix its origin point at a corner of the HRIS.
Its $x$ and $y$ axes are along the directions of the HRIS elements while its $z$ axis can be determined by right-hand rule.
Denote the locations of the BS and the HRIS as $\mathbf p_{B}$ and $\mathbf p_{R}$, respectively,  and define $\mathbf s_B $ as the direction vector of the ULA at BS. Note that $\mathbf p_{R} = (0,0,0)^T$.
Thus, the distance between BS and the HRIS can be represented as $d^{BR} = |\mathbf p_B|$, the AOA at BS is $\theta_{B} = \arccos \frac{\mathbf s_B^T \mathbf p_B}{|\mathbf s_B| \cdot |\mathbf p_B|}$, and the AODs at the HRIS are:
\begin{align}
\phi_{r} &= \arccos \frac{(\mathbf p_B - \frac{\mathbf p_B^T \mathbf e_{z}^R}{|\mathbf e_{z}^R|^2} \mathbf e_{z}^R)^T \mathbf e_{x}^R}{|\mathbf p_B - \frac{\mathbf p_B^T \mathbf e_{z}^R}{|\mathbf e_{z}^R|^2} \mathbf e_{z}^R| \cdot |\mathbf e_{x}^R|},
\\
\psi_{r} &= \pi - \arccos \frac{\mathbf p_B^T \mathbf e_{z}^R}{|\mathbf p_B| \cdot |\mathbf e_{z}^R|},
\end{align}
where $\mathbf e_{x}^R$ and $\mathbf e_{z}^R$ are the unit vectors along the $x$ and $z$ axes in $\mathcal C^R$, respectively.
Then, the combining vector at BS can be designed as $\mathbf r = \frac{1}{N_b} \mathbf a_{\text{B}}(\theta_B)$.

\section{Input-Output Relationship for the HRIS aided OTFS System}

\subsection{HRIS Received Signal Model}

During the ${n_B}$-th time block of the preamble, the HRIS received signal at the $n_T$-th time slot can be represented as
\begin{align}\label{HRIS_receiv_nBnT}
\mathbf y^{RIS}_{n_B}(n_T)
=& \sum_{p=1}^{P} h^{\text{UR}}_{p} e^{\jmath 2 \pi \nu^{\text{UR}}_{p} (((n_B-1) N_T + n_T) T_s - \tau^{\text{UR}}_{p})}
\notag \\
&\times [\mathbf a_R(\phi_p, \psi_p)]_{\mathbf f^a_{n_B}}
\mathbf t(n_T- \tau^{\text{UR}}_{p}/T_s) + \mathbf w_{n_B,n_T}^{RIS}
\notag \\
=& \sum_{p=1}^{P} \overline{h}_p^{\text{UR}} e^{\jmath 2\pi \nu^{\text{UR}}_{p} ((n_B-1) N_T + n_T)T_s}
t_{(n_T - {\tau^{\text{UR}}_{p}}/{T_s})_{N_T}}
\notag \\
&\times [\mathbf a_R(\phi_p, \psi_p)]_{\mathbf f^a_{n_B}} + \mathbf w_{n_B,n_T}^{RIS},
\end{align}
where $n_T \!=\! 0,1,\ldots,N_T-\!1$,
$\overline{h}_p^{\text{UR}} \!=\! h^{\text{UR}}_{p}e^{-\jmath 2 \pi \nu^{\text{UR}}_{p} \tau^{\text{UR}}_{p}} $,
$\mathbf f^a_{n_B}(n_{RF}) \!=\! (n_B-\!1)N_{RF}+\! n_{RF}$ if $n_B \!=\! 1,2,\ldots, \frac{N_x}{N_{RF}}$,
and $\mathbf f^a_{n_B}(n_{RF}) \!=\! ((n_B -\!1) N_{RF} -\! N_x +\! n_{RF} -\!1)N_x +\!1$ if $n_B \!=\! \frac{N_x}{N_{RF}} +\! 1, \frac{N_x}{N_{RF}} +\! 2, \ldots, \frac{2N_x}{N_{RF}}$.
The subscript $(x)_{N_T}$ of $t$ denotes the circular shift of $x$ with respect to $N_T$.
Besides, $ \mathbf w_{n_B,n_T}^{RIS}$ is the noise vector.
Note that $n_{RF} = 1,2,\ldots,N_{RF}$.
To guarantee estimation for the channel gain of any possible delay taps, it is assured that $N_T \ge \frac{\tau_{max}^{\text{UR}}}{T_s} $, where $\tau_{max}^{\text{UR}}$ is the maximum possible channel delay between the user and the HRIS.

Define the Doppler phase shift vector as $\mathbf v (\nu_p^{\text{UR}}) = [1, e^{\jmath 2\pi \nu_p^{\text{UR}} T_s}, \ldots, e^{\jmath 2\pi \nu_p^{\text{UR}} (N_T-1) T_s}]^T$,
and define the reorganized training vector $\mathbf t_t(\tau_{p}^{\text{UR}}) = [t_{(0 - \tau_p^{\text{UR}}/T_s)_{N_T}}, \ldots, t_{(N_T-1 - \tau_p^{\text{UR}}/T_s)_{N_T}}]^T$.
Besides, define the global index for the first time slot of the $n_B$-th time block as $f_{n_B} = (n_B-1) N_T$,
the received signal during the $n_B$-th time block can be written as
\begin{align}\label{RIS_receive_nB}
\mathbf y^{RIS}_{n_B}
=& \sum_{p=1}^{P}
\overline{h}_p^{\text{UR}}
e^{\jmath 2\pi \nu_p^{\text{UR}} f_{n_B} T_s} \big( \mathbf v (\nu_p^{\text{UR}}) \odot \mathbf t_t(\tau_{p}^{\text{UR}}) \big)
\notag \\
&\otimes
[\mathbf a_R(\phi_p, \psi_p)]_{\mathbf f^a_{n_B}} + \mathbf w_{{n_B}}^{RIS}.
\end{align}

Furthermore, define a mixed vector $\mathbf b_R(\nu^{\text{UR}}_{p}, \phi_p, \psi_p) = ([e^{\jmath 2\pi \nu_p^{\text{UR}} f_{1} T_s}, \ldots, e^{\jmath 2\pi \nu_p^{\text{UR}} f_{N_B} T_s} ]^T
\otimes
\mathbf 1_{N_{RF}})
\odot
[[\mathbf a_R(\phi_p, \psi_p)]_{\mathbf f^a_{1}}^T, \ldots, [\mathbf a_R(\phi_p, \psi_p)]_{\mathbf f^a_{N_B}}^T]^T
$,
we can collect the received signal of all $N_B$ time blocks, and rearrange them as
\begin{align}\label{RIS_receive}
\mathbf y^{RIS}
\!\!\!=\!& \! \sum_{p=1}^{P}
\!\overline{h}_p^{\text{UR}}
\!\!\big(\! \mathbf v (\nu_p^{\text{UR}}\!) \!\odot\! \mathbf t_t(\tau_{p}^{\text{UR}}) \! \big) \!\!\otimes\!\!
\mathbf b_R(\nu^{\text{UR}}_{p}\!\!, \phi_p, \psi_p\!)
\!\!+\! \mathbf w^{RIS}.
\end{align}


\subsection{BS Received Signal Model over Delay-Doppler Domain}

For the sake of the subsequent derivation, we first define a grid set in time-frequency domain as
$
\mathcal{G}^{\text{TF}} \!=\!\left\{ \left. \left(n T, \!m \Delta f\right) \ \right| \ n\!=\!0,1,\cdots,\!N\!-\!1 \ , \!\ m\!=\!0,1,\cdots,\!M\!-\!1 \right\}
$,
where $T$ and $\Delta f$ are sampling intervals along the time and frequency dimensions, respectively.
$N$ is the number of time slots, and $M$ is the number of subcarriers.
Correspondingly, the grid set in delay-Doppler domain is defined as
$
\mathcal{G}^{\text{DD}}\!\!=\!\!\left\{\!\left. \left(\frac{k}{N T}, \frac{l}{M \Delta f} \!\right) \ \!\right| \!\ k\!\!=\!\!-\frac{N}{2},\cdots,\frac{N}{2}\!-\!1, \ l\!=\!0,1,\cdots,M\!-\!1 \!\right\}
$,
where $\frac{1}{M\Delta f}$ and $\frac{1}{NT}$ are sampling intervals along the delay and Doppler dimensions, respectively.
Note that $T_s = 1/ M \Delta f$.

We arrange a set of $N\times M$ information symbols in delay-Doppler domain on the grid $\mathcal{G}^{\text{DD}}$ of each OTFS block for the user.
Within $\mathcal{G}^{\text{DD}}$, we place $N_P \times M_P$ pilot symbols and $(N-N_P) \times M_P$ guard grids, while other remaining grids are all filled with data symbols.
Without loss of generality, we assume that the index set of the transmitted pilot grids for each OTFS block is $\mathcal C_P = \{(k,l)| k\in [-N/2,N_P-N/2-1], l\in (0, [0 + M_{P} -1]_M)\}$, and the index set of its receiving grids is defined as $\mathcal D_P = \{(k,l)|k\in [-N/2, N/2-1], l\in [0, M_{P} + l_{\tau_{{max}}} -1]_M \}$ for further use.
In order to send the delay-Doppler symbols $X^{\text{DD}}_{k,l} $, the user first applies the inverse symplectic finite Fourier transform
to convert the symbols into time-frequency domain symbols as
$
X^{\text{TF}}_{n,m} = \frac{1}{\sqrt{NM}}
\sum\limits_{k=-\frac{N}{2}}^{\frac{N}{2}-1} \sum\limits_{l=0}^{M-1} X^{\text{DD}}_{k,l} e^{\jmath 2 \pi\left(\frac{nk}{N}-\frac{m l}{M}\right)}
$,
where $n = 0, \ldots, N-1 $ and $m = 0, \ldots, M-1 $.
Next, the time-frequency domain signal is transformed into a time domain signal $s(t)$ through Heisenberg transform as
$
s(t)=\sum\limits_{n=0}^{N-1} \sum\limits_{m=0}^{M-1} X^{\text{TF}}_{n,m} g_{tx}(t-n T) e^{\jmath 2 \pi m \Delta f(t-nT)}
$,
where the transmit pulse $g_{tx}(t)$ is a rectangular function with duration $T$.
After $s(t)$ is transmitted from the user, the noiseless impinging signal $\mathbf{y}^{\text{R}}(t)\in\mathbb{C}^{N_r\times 1}$ received at the HRIS is given by
\begin{align}
\mathbf{y}^{\text{R}}(t)
&= \sum_{p=1}^{P} h^{\text{UR}}_{p} e^{\jmath 2 \pi \nu^{\text{UR}}_{p} (t - \tau^{\text{UR}}_{p})} \mathbf{a}_{\text{R}} (\phi_p,\psi_p)
s(t- \tau^{\text{UR}}_{p}).
\end{align}
Let us denote the HRIS phase shift vector at time $t$ as
$\boldsymbol{\omega}(t)\triangleq[e^{\jmath\omega_1(t)},\cdots,e^{\jmath\omega_{N_r}(t)}]^T$, where $\omega_{r}(t)\in[0,2\pi), r\in\{0,1,\cdots,N_r-1\}$ represents the phase shift of the $r$-th HRIS element.
Then, the operation of HRIS is described by the diagonal matrix $\boldsymbol{\Omega}(t)=\text{diag}\{\boldsymbol{\omega}(t)\}\in\mathbb{C}^{N_r\times N_r}$, which will be designed in the next subsection for better performance.
As we aim to achieve the channel estimation and data detection simultaneously, the designed HRIS phase shift matrix $\boldsymbol{\Omega}(t)$ is constant within one OTFS block.
Without loss of generality, we omit the time index $t$ of both $\boldsymbol{\omega}$ and $\boldsymbol{\Omega}$.
Then, the combined received signal at the BS can be obtained as
\begin{align}
y^{\text{B}}(t)
\!=& \mathbf {r}^{H} \sum_{i_B=0}^{I_B-1} \mathbf H_{i_B}^{\text{RB}}(t) \boldsymbol{\Omega} \mathbf{y}^{R}(t- i_BT_s) + \mathbf w^B (t)
\notag\\
=& h^{\text{RB}}
\mathbf{a}^H_{\text{R}} (\phi_{\text{r}}, \psi_{\text{r}}) \boldsymbol{\Omega}
\sum_{p=1}^{P} h^{\text{UR}}_{p} e^{\jmath 2 \pi \nu^{\text{UR}}_{p} (t- \tau^{\text{RB}} - \tau^{\text{UR}}_{p})} \mathbf{a}_{\text{R}} (\phi_p,\psi_p)
\notag \\
&\times s(t - \tau^{\text{RB}} - \tau^{\text{UR}}_{p}) + \mathbf w^B (t),
\label{eq:re_user}
\end{align}
where $\mathbf w^B (t)$ is the additive white Gaussian noise (AWGN).
We denote $\tau_p = \tau_p^{\text{UR}}+\tau^{\text{RB}}$, $\nu_{p}=\nu^{\text{UR}}_{p}$, $h_p = h^{RB} h^{UR}_p$, $p\in\{1,\cdots,P\}$,
and denote the combined equivalent channel gain of the $p$-th scattering path as
$
\widetilde{h}_p = h_{p}
\mathbf{a}^H_{\text{R}} (\phi_{\text{r}}, \psi_{\text{r}})
\boldsymbol{\Omega}
\mathbf{a}_{\text{R}} (\phi_p,\psi_p)$.
Then \eqref{eq:re_user} can be modified as
\begin{align}
y^{\text{B}}(t)
&= \sum_{p=1}^{P}
\sum_{n=0}^{N-1} \sum_{m=0}^{M-1}
\widetilde{h}_p  e^{\jmath 2 \pi \nu_{p} (t- \tau_{p})}
X^{\text{TF}}_{n,m} g_{tx}(t - \tau_{p} - n T)
\notag\\
&\times e^{\jmath 2 \pi m \Delta f(t-\tau_{p}-nT)} + \mathbf w^B (t).
\end{align}
Furthermore, $y^{\text{B}}(t)$ is transformed back to the time-frequency domain through inverse of Heisenberg transform with a rectangular pulse.
Similar with the operations in \cite{OTFS2},
by sampling over the time-frequency domain and implementing the SFFT to the sampled signal,
we can represent the received signal in delay-Doppler domain as
%
%
\begin{align}\label{DD_y1}
Y^{\mathrm{DD}}_{k,l}
=&\! \sum_{p=1}^{P} \!\sum_{q=0}^{N-1}
\widetilde{h}_{p}
\gamma(k,l,l_{\tau_p},q,k_{\nu_p}, \beta_{\nu_p})
\notag \\
&\times X^{\text{DD}}_{[k-k_{\nu_p} \!+ q]_N- \frac{N}{2}, [l-\!l_{\tau_p}]_M}
+\! W^{\text{DD}}_{k,l},
\end{align}
where $W^{\text{DD}}_{k,l} \sim \mathcal {CN}(0, \sigma_n^2)$ is the AWGN in delay-Doppler domain, $l_{\tau_p}$ is the tap of $\tau_p$ which can be represented as
$\l_{\tau_p} = \tau_p M \Delta f$.
$k_{\nu_p}$ and $\beta_{\nu_{p}} \in [-0.5, 0.5)$ are respectively the integer tap and its related fractional bias of the Doppler frequency shift $\nu_p$, which are calculated as $\nu_p = (k_{\nu_p} + \beta_{\nu_p})/NT$.
In addition,
\begin{align}
&\gamma(k,l,l_{\tau_p},q,k_{\nu_p}, \beta_{\nu_p})
\notag \\
=\!\!&\left\{\begin{array}{ll}
\!\!\!\!\frac{1}{N} \xi(l, \!l_{\tau_p}, \!k_{\nu_p}, \!\beta_{\nu_p}\!)
\theta\!\left(q, \!\beta_{\nu_{p}}\!\right)\!,
\!\!\!\!\!& l_{\tau_p} \!\!\leq \!l\!\! < \!M, \\
\!\!\!\!\frac{1}{N} \xi(l, \!l_{\tau_p}, \!k_{\nu_p}, \!\beta_{\nu_p}\!)
\theta\!\left(q, \!\beta_{\nu_{p}}\!\right)
\!\phi\left(k, q, k_{\nu_{p}}\right)\!,
\!\!\!\!\!& 0 \!\!\leq \!l\!\! < \!l_{\tau_p},
\end{array}\right.
\end{align}
where $\xi(l, l_{\tau_p}, k_{\nu_p}, \beta_{\nu_p})
\!\!=\!\! e^{j 2 \pi ( \frac{l- l_{\tau_p}}{M} ) ( \frac{k_{\nu_{p}} + \beta_{\nu_{p}}}{N} )}$, $\theta(q, \beta_{\nu_p})
\!\!=\!\! \frac{e^{-\!j 2\pi (\frac{N}{2}-q - \beta_{\nu_p})}-\!1}{e^{-\!j \frac{2\pi}{N} (\frac{N}{2}-q - \beta_{\nu_p})}-\!1}$, and $\phi(k,q,k_{\nu_p})
\!\!=\!\! e^{-\!j 2\pi \frac{[k-k_{\nu_p} +q]_N}{N}}$.

Define $\boldsymbol \gamma_{k,l,p} \!=\! [\gamma_{k,l,1,0}, \ldots \gamma_{k,l,1,N\!-\!1}]^T$ and
$\mathbf x_{k,l,p}^{\text{DD}} \!=\! [x_{k,l,p,0}^{\text{DD}}, \ldots, x_{k,l,p,N\!-\!1}^{\text{DD}}]^T$,
where $\gamma_{k,l,p,q} \!\triangleq\! \gamma(k,l,l_{\tau_{p}},q,k_{\nu_{p}}, \beta_{\nu_{p}})$
and $x_{k,l,p,q}^{\text{DD}} = X^{\text{DD}}_{[k-k_{\nu_p} + q]_N- \frac{N}{2}, [l-l_{\tau_p}]_M}$.
Then we denote ${\mathbf z}_{k,l} = [{z}_{k,l}[1], \ldots, {z}_{k,l}[P]]^T$,
where ${z}_{k,l}[p] = \boldsymbol \gamma_{k,l,p}^T \mathbf x_{k,l,p}^{\text{DD}}$.
Besides, define $\widetilde{\mathbf h} = [\widetilde{h}_{1}, \widetilde{h}_{2}, \ldots, \widetilde{h}_{P}]^T$, and thus \eqref{DD_y1} can be rewritten as
$
Y^{\mathrm{DD}}_{k,l}
= {\mathbf z}_{k,l}^T \widetilde{\mathbf h} + W^{\text{DD}}_{k,l}
$.
Furthermore, define $\mathbf y^{\text{DD}} \!=\! [Y^{\mathrm{\text{DD}}}_{-\!N/2,0}, \ldots, Y^{\mathrm{\text{DD}}}_{N/2-1,M-\!1}]^T \!\in\! \mathbb C^{MN \times\! 1}$,
${\mathbf Z} \!=\! [{\mathbf z}_{-\!N/2, 0}, \ldots, {\mathbf z}_{N/2-\!1,M-\!1}] $
and $\mathbf w^{\text{DD}} =  \text{vec}(\mathbf W^{\text{DD}})$,
the whole delay-Doppler domain received signal can be finally represented as
\begin{align}\label{representation_for_FG}
\mathbf y^{\text{DD}}
&= {\mathbf Z}^T \widetilde{\mathbf h} + \mathbf w^{\text{DD}}.
\end{align}

\subsection{HRIS Beamforming Design and Calibration}

From \eqref{RIS_receive},
it can be checked that the rearranged received signal at the HRIS is the summation of multiple components, thus we can resort to Newtonized orthogonal match pursuit (NOMP) algorithm to estimate the parameters at the HRIS \cite{OTFS_li}.
Note that the AOAs' variation between two preambles is very small, so the angle searching range can be set very small, and the estimated AOA of the last preamble can be regard as the searching center.
Then, the delay $\tau_{p}^{\text{UR}}$, the Doppler frequency shift $\nu^{\text{UR}}_{p}$, the AOAs $\{\phi_p, \psi_p\}$, and the channel gain ${\overline{h}}_p^{\text{UR}}$ for each paths between the user and the HRIS will be obtained at the HRIS.
Thus, ${h}^{\text{UR}}_{p}$ can be derived as ${h}^{\text{UR}}_{p} = {\overline{h}}_p^{\text{UR}} e^{\jmath 2 \pi \nu^{\text{UR}}_{p} \tau^{\text{UR}}_{p}} $.
Assume that each HRIS elements induce independent phase shifts on the incident signals.
Since data symbols occupy almost all the grids over the delay-Doppler domain, it can be approximated that the mean power of all the transmitted symbols is the same with the power of data symbol $\sigma_D^2$.
According to \eqref{DD_y1}, the received signal-to-noise ratio (SNR) of $Y^{\text{DD}}_{k,l} $ can be represented as
\begin{align}\label{y1_SNR_kl}
\rho_{k,l}
=&
\frac{\sigma_D^2}
{\sigma_n^2}
|h^{\text{RB}}|^2 \sum\limits_{p=1}^{P} |{h}_p^{\text{UR}}|^2
|\mathbf{a}^H_{\text{R}} (\phi_{\text{r}}, \psi_{\text{r}})
\boldsymbol{\Omega}
\mathbf{a}_{\text{R}} (\phi_p,\psi_p)|^2
\notag\\
&\times \sum\limits_{q=0}^{N-1}
|\gamma(k,l,l_{\tau_p},q,k_{\nu_p}, \beta_{\nu_p})|^2.
\end{align}
Then the optimization problem for the phase shift matrix $\mathbf \Omega$ design can be formulated as
\begin{align}
\text{(P1):} \max_{\mathbf \Omega}\ \  &
\sum_{k = -\frac{N}{2}}^{\frac{N}{2}-1} \sum_{l = 0}^{M-1}
\log_2 (1+\rho_{k,l}),
\notag \\
\text{s.t.}\ \  &|\omega_r| = 1, \forall r = 1,2,\ldots,N_r. \label{normalized_condition}
\end{align}
Although $h_p^{\text{UR}}$ for $p = 1,2,\ldots,P$ have been estimated,
$h^{\text{RB}}$ is not known for the beamforming design.
Fortunately, $h^{\text{RB}}$ is the same for each scattering path at the UE side, and is constant within the duration of one OTFS block.
Hence, the influence of $h^{\text{RB}}$ can be normalized in the HRIS beamforming design.
Moreover, with the monotonicity of the logarithmic function and the independence of noise and the transmitted symbols, (P1) can be equivalently transformed as
\begin{align}
\text{(P2):} \max_{\boldsymbol \omega} \ \ &
\sum\limits_{p=1}^{P} |{h}_p^{\text{UR}}|^2
|\mathbf{a}^H_{\text{R}} (\phi_{\text{r}}, \psi_{\text{r}})
\odot
\mathbf{a}_{\text{R}}^T (\phi_p,\psi_p)
\boldsymbol{\omega}|^2
\notag \\
&\times
\sum\limits_{(k,l) \in \mathcal{C} \setminus \mathcal C_P}
\sum\limits_{q=0}^{N-1}
|\gamma(k,l,l_{\tau_p},q,k_{\nu_p}, \beta_{\nu_p})|^2
, \notag \\
\text{s.t.}\ \  &|\omega_{n_r}| = 1, \forall r = 1,2,\ldots,N_r.
\label{normalized_condition2}
\end{align}
Furthermore, by defining
$
|S_p|^2 =
\sum\limits_{(k,l) \in \mathcal{C} \setminus \mathcal C_P}
\sum\limits_{q=0}^{N-1}
|\gamma(k,l,l_{\tau_p},q,k_{\nu_p}, \beta_{\nu_p})|^2
$,
the optimal solution without the constraint can be readily obtained as
$
\boldsymbol{\omega}^{\star} =
\sum\limits_{p=1}^{P}
\left(|S_p| |{h}_p^{\text{UR}}| \mathbf{a}^H_{\text{R}} (\phi_{\text{r}}, \psi_{\text{r}})
\odot
\mathbf{a}_{\text{R}}^T (\phi_p,\psi_p)\right)^H
$.
Due to the normalization constraint in \eqref{normalized_condition2}, we normalize each element in $\boldsymbol{\omega}^{\star}$ as
\begin{align}\label{beamforming_design}
{\omega}^{\star}_{n_r} =
\frac{\sum\limits_{p=1}^{P}
(|S_p| |{h}_p^{\text{UR}}|)
[\mathbf{a}_{\text{R}} (\phi_{\text{r}}, \psi_{\text{r}})]_{n_r}
[\mathbf{a}^{\ast}_{\text{R}} (\phi_p,\psi_p)]_{n_r}}
{\left|\sum\limits_{p=1}^{P}
(|S_p| |{h}_p^{\text{UR}}|)
[\mathbf{a}_{\text{R}} (\phi_{\text{r}}, \psi_{\text{r}})]_{n_r}
[\mathbf{a}^{\ast}_{\text{R}} (\phi_p,\psi_p)]_{n_r}\right|}.
\end{align}

Within the duration of each short pilot sequence before the accompanying OTFS block, we assume that the first $N_{RF}$ elements of the first column on the HRIS are sampled without loss of generality.
According to \eqref{RIS_receive_nB},
the sampled received signal at the HRIS can be represented as
\begin{align}
\overline{\mathbf y}^{RIS} =& \mathbf J_t
\overline{\mathbf h}^{\text{UR}} +  \mathbf w^{RIS}
,
\end{align}
where $[\mathbf J_t]_{:,p} = e^{\jmath 2\pi \nu_p^{\text{UR}} T_s} \big( \mathbf v (\nu_p^{\text{UR}}) \odot \mathbf t_t(\tau_{p}^{\text{UR}}) \big)
\otimes
[\mathbf a_R(\phi_p, \psi_p)]_{1:N_F}$, and $ \overline{\mathbf{h}}^{UR} = [\overline{{h}}^{UR}_1, \overline{{h}}^{UR}_2, ..., \overline{{h}}^{UR}_P]^T $.

Although the user is moving, the scattering environment can be regarded as constant within several OTFS blocks due to the far distance between the user and the HRIS.
For example, within a OTFS system, we set the number of subcarriers as $N = 256$, the number of time slots as $M = 16$, and the sampling rate as $1/ T_s = 20$ MHz.
Then, if the user moves at the speed $360$ km/h and the distance between the user and the RIS is $200$  m, the maximal angle change within $20$ OTFS blocks is about $0.12^{\circ}$.
Hence, the channel parameters except the channel gain can be seen as the same.
Therefore, $\mathbf J_t$ is constant within a long duration.
Through a simple least square (LS) method, the channel gain can be updated as
$\widehat{\mathbf h}^{\text{UR}}
=
e^{\jmath 2 \pi \nu^{\text{UR}}_{p} \tau^{\text{UR}}_{p}}
\mathbf J_t^H(\mathbf J_t \mathbf J_t^H)^{-1} \mathbf y^{RIS}$.
Then, $\boldsymbol{\omega}^{\star}$ can be updated by substituting the newly estimated ${h}_p^{\text{UR}} $ into \eqref{beamforming_design}.

\section{Joint Channel Estimation and Data Detection over delay-Doppler Domain}

Define a maximum delay spread $\tau_{{max}}$ of the cascaded channel, and each delay tap within the range is related with one possible scattering path.
Hence, the $P$ actual channel paths can be mapped in part of the delay domains, i.e., $l_{\tau_p} \in [0,\tau_{max}]$.
Then $\widetilde{\mathbf h}$ becomes a $P_{max} \times 1$ sparse vector, where $P_{max} = \tau_{max}/T_s+1$.
Note that there are $P$ non-zeros elements in $\widetilde{\mathbf h}$.
The probability distribution function (PDF) $p(\widetilde{h}_{p})$ for the $p$-th element in $\widetilde{\mathbf h}$ is assumed to be a Bernoulli-Gaussian distribution as
$
p(\widetilde{h}_{p})
= (1 - \alpha)\delta(\widetilde{h}_{p}) + \alpha \mathcal{CN}(\widetilde{h}_{p}; 0, \lambda_{p})
$,
where $\alpha$ is the delay-domain channel sparsity factor, and $\lambda_{p}$ is the prior variance of the non-zero part.
Define $\mathbf \Lambda = \text{diag}(\lambda_{1}, \ldots, \lambda_{P_{max}}) \in \mathbb{C}^{P_{max} \times P_{max}}$ for further use.
Since it has been illustrated that the scattering environment can be seen as constant within dozens of OTFS blocks.
Hence, it can be assumed that the PDFs of $\widetilde{\mathbf h}$ for dozens of OTFS blocks are the same.
Thus, we adopt $N_O$ OTFS blocks for the accuracy of channel characteristic estimation.
For notation simplicity, we will only show the block index $n_o$ in the estimation procedure of $\alpha$ and $\mathbf \Lambda$.

It is worth noting that the index set of transmitted signal grids for the receive index $\mathcal D_p$ is $\mathcal C_{\mathcal D_P} \!=\! \{(k,l)| k\!\in\! [-N/2, N/2\!-\!1], l\!\in\! [[-l_{\tau_{{max}}}]_M, [M_{P} \!+ \!l_{\tau_{{max}}} \!-\!1]_M]\}$.
Hence, the channel gain estimation will be influenced by the data symbols placed within the index set $\{(k,l)|k \!\in\! [-N/2, N/2\!-\!1], l \!\in\! \{[M_P-1, M_{P} \!+\! l_{\tau_{{max}}} \!-\! 1], [M \!-\! l_{\tau_{{max}}}, M \!-\! 1]\}\}$.
Besides, to combat with the interference between pilot and data over the delay-Doppler domain, we adopt a power gap factor $\Delta$, i.e., $\sigma_P^2 = \sqrt{10^{(\Delta/10)}} \sigma_D^2$, where $\sigma_P^2$ is the average power of pilot symbols.

\begin{figure*}[htbp]
 \centering
 \includegraphics[width=120mm]{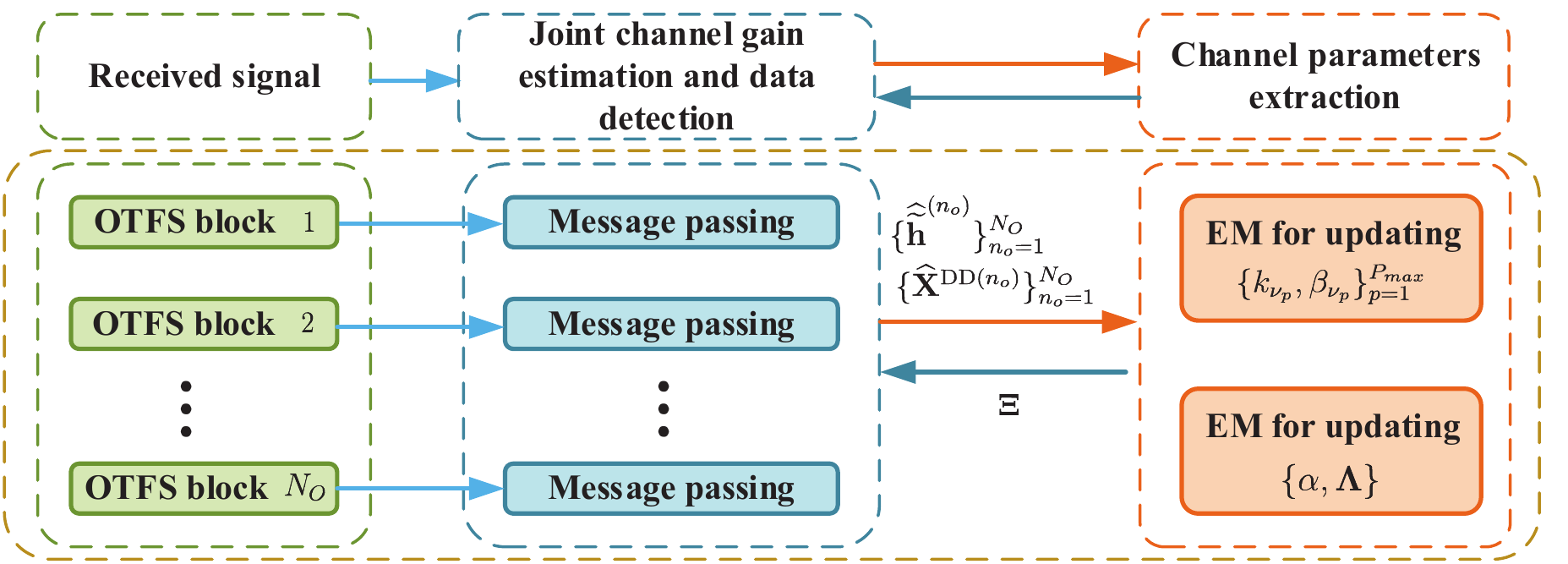}
 \caption{Block diagram of the proposed joint channel estimation and data detection scheme.}
 \label{Alg_framework}
\end{figure*}

Since the received signal at BS is highly correlated with the channel characteristic, the Doppler shifts on each delay tap should be updated within the process of the joint channel estimation and data detection.
Our idea is to simultaneously obtain the CSI and data by iteratively implementing MP algorithm and EM algorithm, which is illustrated in  \figurename{ \ref{Alg_framework}}.
MP is to acquire the unknown data symbols and the a posteriori distribution of the equivalent channel gain simultaneously, while EM algorithm is to update the channel parameters $\boldsymbol \Xi = \{ \alpha, \mathbf \Lambda, \{k_{\nu_p}, \beta_{\nu_p}\}_{p=1}^{P_{max}}\}$ to support the MP process.
Similarly, the posterior statistics produced by MP will also benefit the EM procedure.
Note that there are iterative operations not only between MP and EM procedure, but also within both the two algorithm frameworks.
This can raise the channel estimation accuracy at the first iteration, and improve the convergence of the algorithm.

\subsection{Message Passing based Joint Equivalent Channel Gain Estimation and Data Detection}


\subsubsection{\bf Factor Graph Representation}

Let $\mathcal M$ be the constellation set of each symbol $X^{\text{DD}}_{k,l}$, and $|\mathcal M|$ be the size of $\mathcal M$. For each $X^{\text{DD}}_{k,l}$, it can be assumed that
$p(X^{\text{DD}}_{k,l} \!=\! c^s) \!=\! \frac{1}{|\mathcal M|}$ for any $c^s \!\in \! \mathcal M$ if $X^{DD}_{k,l}$ is a data symbol, while $p(X^{\text{DD}}_{k,l} \!=\! c^s) \!=\! 1$ if $X^{\text{DD}}_{k,l}$ is a pilot symbol, where $c^s$ is a value of $X^{\text{DD}}_{k,l}$.
Besides, $\mathbf y^{\text{DD}} \sim \mathcal {CN}(\mathbf Z^T \widetilde{\mathbf h}, \sigma_n^2)$.
In addition, $\widetilde{h}_p$ for $p = 1, \ldots, P_{max}$ is i.i.d. and is distributed as $p(\widetilde{h}_p)$.
With \eqref{representation_for_FG}, the joint distribution of $\mathbf y^{\text{DD}}$, $\mathbf X^{\text{DD}}$ and $\widetilde{\mathbf h}$ can be expressed as
\begin{align}\label{P_representation_for_FG}
p(\mathbf y^{\text{DD}}, \mathbf Z, \widetilde{\mathbf h})
=& \prod_{(k,l) = (-\frac{N}{2},0)}^{(\frac{N}{2}-1, M-1)} p(y^{\text{DD}}_{k,l} \big| \mathbf z_{k,l}, \widetilde{\mathbf h})
\prod_{p = 1}^{P_{max}} p(\widetilde{h}_p)
\notag \\
& \times
\prod_{(k,l) = (-\frac{N}{2},0)}^{(\frac{N}{2}-1, M-1)} \prod_{p = 1}^{P_{max}} p(z_{k,l}[p], \mathbf x_{k,l,p}^{\text{DD}})
,
\end{align}
where
\begin{align}
p(z_{k,l}[p], \mathbf x_{k,l,p}^{\text{DD}})
\!\!=\!\! p(z_{k,l}[p] \big| \mathbf x_{k,l,p}^{\text{DD}})
\!\!\prod\limits_{q = -\frac{N}{2}}^{\frac{N}{2}-1} \!\!p(X^{\text{DD}}_{q,[l-l_{\tau_{p}}]_M})
.
\end{align}

\begin{figure}[htbp]
	\centering
	\includegraphics[width=85mm]{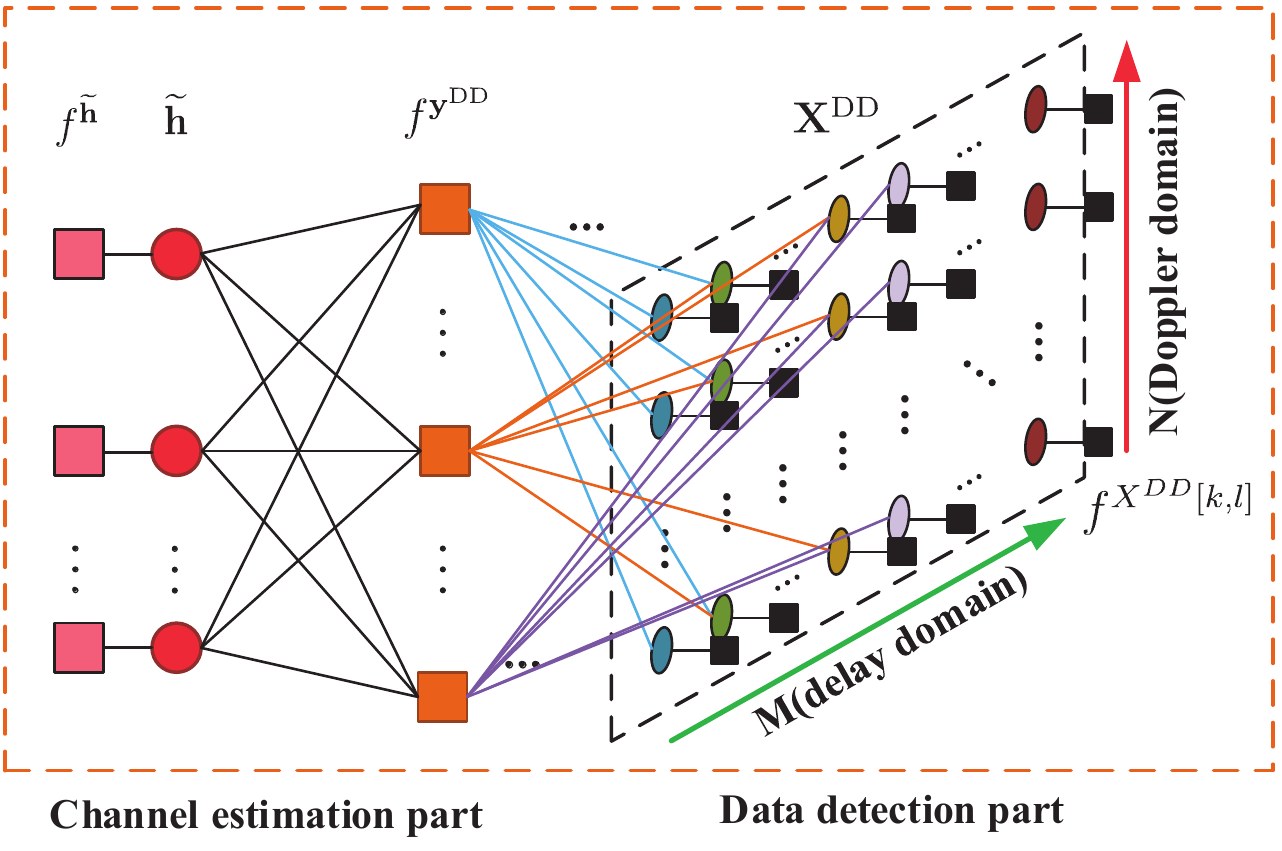}
	\caption{Factor graph of the joint channel estimation and data detection.}
	\label{factor_graph_joint}
\end{figure}

By treating $\{\widetilde{h}_{p}\}_{p=1}^{P_{max}}$ and $\{X^{\text{DD}}_{k,l}\}_{(k,l) = (-\frac{N}{2}, 0)}^{(\frac{N}{2}-1,M-1)}$
as variable nodes, the factor graph based representation of \eqref{P_representation_for_FG} is depicted in \figurename{ \ref{factor_graph_joint}}.
We divide the factor graph into two parts.
The left part focuses on the equivalent channel gain estimation, while the right part is for data detection.
The interaction between the two parts will be implemented through a turbo framework \cite{TMP}.

Focusing on the right part of the factor graph, we denote the set of the receiving grids of one transmitted symbol $x_{k,l,p,q}$ as $\mathcal G_{y^{k,l}}^{x^{p,q}} \!=\! \{(k',l')| k' \!\in \! \{-\frac{N}{2}, \ldots, \frac{N}{2}\!-\!1\}, |l' \!-\! [l\!-\!l_{\tau_p}]_M| \!\leq \! l_{\tau_{{max}}} \}$.
Thus, its corresponding mapping index set can be defined as $\mathcal J_{y^{k,l}}^{x^{p,q}} \!=\!\! \{(k',\!l',\!p',\!q') | (k',\!l') \!\in\! \mathcal G_{y^{k,l}}^{x^{p,q}}, l'\!-\!p' \!=\! l\!-p, k'\!-k_{\nu_{p'}} \!+ q' \!=\! k - k_{\nu_{p}} \!+ q\}$.
Besides, the set of transmitted symbol grids corresponding to $y^{\text{DD}}_{k,l}$ is defined as $\mathcal G_{x}^{y^{k,l}} \!=\! \{(p,\!q)| p \!\in\! \{1,\!2,\ldots, \!P_{max}\!\}, q \!\in \!\{0,\!1, \ldots, \!N\!-\!1\}\}$.
In addition, we define $\overline{\mathcal J}_{y^{k,l}}^{x^{p,q}} \!=\! \{(k',\!l',\!p',\!q')|  (k',\!l',\!p',\!q') \!\in\! \mathcal J_{y^{k,l}}^{x^{p,q}} \!\setminus\! (k,\!l,\!p,\!q) \}$ and $\overline{\mathcal G}_{x^{p,q}}^{y^{k,l}} \!=\! \{(p',\!q')| (p',\!q') \!\in \!\mathcal G_{x}^{y^{k,l}} \!\setminus\! (p,\!q) \}$ for further use.
On the other hand, the left part of the factor graph also has similar structure but with fully connection between the elements of $\widetilde{\mathbf h}_{p}$ and $\mathbf y^{\text{DD}}$.
Therefore, we can separately resort to the idea of belief propagation (BP) for both the two parts.

\subsubsection{\textbf{Right part for data detection}}
Define $\bar{y}_{k,l} \triangleq \sum\limits_{p=1}^{P_{max}} \widetilde{h}_p \boldsymbol \gamma_{k,l,p}^T \mathbf x_{k,l,p}^{\text{DD}}$
and $f^{y_{k,l}} = \mathcal{CN}\left(y_{k,l}^{\text{DD}}; \bar{y}_{k,l}, \sigma_n^2 \right)$.
For simplicity, we will omit the superscript of $y_{k,l}^{\text{DD}}$ and $ \mathbf x_{k,l,p}^{\text{DD}}$ in the following.
Then, the message from function node $f^{y_{k,l}}$ to variable node $x_{k,l,p,q}$ can be represented as
\begin{align}
&\vartheta_{f^{y_{k,l}} \to x_{k,l,p,q} }
\notag\\
&\approx
\!\!\int_{\widetilde{\mathbf h}} \prod_{p'=1}^{P_{max}} \vartheta_{\widetilde{h}_{p'} \to f^{y_{k,l}}}
\int_{\bar{y}_{k,l}}
p\left(y_{k,l} | \bar{y}_{k,l}, \widetilde{\mathbf h}, {\mathbf X}^{DD}\right)
\notag\\
&\quad\times
\mathcal{CN}\left(\bar{y}_{k,l}; \widehat{\bar{y}}_{k,l}, \lambda^{\bar{y}}_{k,l}\right)
\notag\\
&\propto \mathcal {CN} \left(x_{k,l,p,q}; \mu_{f^{y_{k,l}} \to x_{k,l,p,q}}, \eta_{f^{y_{k,l}} \to x_{k,l,p,q}} \right),
\end{align}
where the distribution of $\bar{y}_{k,l}$ given $\widetilde{\mathbf h}$ and $\mathbf X^{DD}$ is the circularly symmetric complex Gaussian (CSCG) distribution with mean $\widehat{\bar{y}}_{k,l}^x$ and variance $\lambda^{\bar{y},x}_{k,l}$ given by
$\widehat{\bar{y}}_{k,l}^x =  \widetilde{h}_p \gamma_{k,l,p,q} x_{k,l,p,q} + \widehat{u}_{p,q}^{y,x}$ and $\lambda^{\bar{y},x}_{k,l} = v_{k,l,p,q}^{u_{y,x}}$,
and
\begin{align}
\widehat{u}_{p,q}^{y,x} \!\!\triangleq&\!\! \sum_{(p',q') \in \overline{\mathcal G}_{x^{p,q}}^{y^{k,l}}}
\mu_{{\widetilde{h}_{p'}} \to f^{y_{k,l}}}
\gamma_{k,l,p',q'} \mu_{x_{k,l,p',q'} \to f^{y_{k,l}}}, \label{part_mean_z}
\\
v_{k,l,p,q}^{u_{y,x}} \!\!\triangleq&\!\! \sum_{(p',q') \in \overline{\mathcal G}_{x^{p,q}}^{y^{k,l}}}
\!\!|\gamma_{k,l,p',q'}|^2\!
\left(
|\mu_{{\widetilde{h}_{p'}} \to f^{y_{k,l}}}|^2 \eta_{x_{k,l,p',q'} \to f^{y_{k,l}}}
 \right.
\notag\\
&+ \!\!\left.
\eta_{{\widetilde{h}_{p'}} \!\to\! f^{y_{k\!,l}}}\!
\big(\!\eta_{x_{k,l,p'\!,q'} \!\to\! f^{y_{k\!,l}}} \!\!+\!\! |\!\mu_{x_{k,l,p'\!\!,q'} \!\to\! f^{y_{k\!,l}}}|^2 \big)\!\!
\right)
.
\label{part_var_z}
\end{align}
$\{\mu_{{\widetilde{h}_{p'}} \to f^{y_{k,l}}}, \eta_{{\widetilde{h}_{p'}} \to f^{y_{k,l}}}\}$ and  $\{\mu_{x_{k,l,p',q'} \to f^{y_{k,l}}}, \eta_{x_{k,l,p',q'} \to f^{y_{k,l}}}\}$ are respectively the mean and variance of the message $\vartheta_{{\widetilde{h}_{p'}} \to f^{y_{k,l}}}$ from variable node ${\widetilde{h}_{p'}}$ to function node $f^{y_{k,l}}$ and the message $\vartheta_{x_{k,l,p',q'} \to f^{y_{k,l}}}$ from variable node $x_{k,l,p',q'}$ to function node $f^{y_{k,l}}$ in the last iteration, which will be defined later.
Note that $\vartheta_{{\widetilde{h}_{p'}} \to f^{y_{k,l}}}$ can be initialized with the priori PDF of ${\widetilde{h}_{p'}}$,
and $\vartheta_{x_{k,l,p',q'} \to f^{y_{k,l}}}$ can be initialized as the Gaussian projection of priori probability for $x_{k,l,p',q'}$ in the first iteration \cite{OTFS_Ge2}.
Besides, according to Central Limit Theorem \cite{CLT}, we can obtain
\begin{align}
\mu_{f^{y_{k,l}} \!\to\! x_{k,l,p,q}}\!\!
&=\!\! \frac{y_{k,l} - \widehat{u}_{p,q}^{y,x}}
{\gamma_{k,l,p,q} \mu_{{\widetilde{h}_{p}} \to f^{y_{k,l}}}},
\\
\eta_{f^{y_{k,l}} \!\to\! x_{k,l,p,q}}\!\!
&=\!\! \frac{v_{k,l,p,q}^{u_{y,x}}
\!\!+\!\! \sigma_n^2 \!\!-\!\! |\!\gamma_{k,l,p,q}\!|^2 |\widehat{r}_{k,l,p,q}\!|^2 \eta_{{\widetilde{h}_{p}} \!\to\! f^{y_{k,l}}}
}
{|\gamma_{k,l,p,q}|^2 \!\left(|\mu_{{\widetilde{h}_{p}} \!\to\! f^{y_{k,l}}}|^2 \!\!+\!\! \eta_{{\widetilde{h}_{p}} \!\to\! f^{y_{k,l}}} \!\right) }.
\end{align}

Combining the messages from function nodes $f^{y_{k',l'}}$ to the variable node $x_{k,l,p,q}$ where $(k',l') \in \mathcal G_{y^{k,l}}^{x^{p,q}}$ and $(k',l') \ne (k,l)$,
we can obtain a Gaussian message with the variance and the mean given respectively by
\begin{align}
\eta_{f^{\mathbf y \setminus y_{k,l}} \to x_{k,l,p,q}}^{sum}
=& \Big( \sum_{(k',l',p',q') \in \overline{\mathcal J}_{y^{k,l}}^{x^{p,q}} }
\frac{1}{\eta_{f^{y_{k',l'}} \to x_{k',l',p',q'}}} \Big)^{-1}, \\
\mu_{f^{\mathbf y \setminus y_{k,l}} \to x_{k,l,p,q}}^{sum}
=& \eta_{f^{\mathbf y \setminus y_{k,l}} \to x_{k,l,p,q}}^{sum}
\notag \\
&\times
\!\!\!\sum_{(k',l',p',q') \in \overline{\mathcal J}_{y^{k,l}}^{x^{p,q}} } \!\!
\frac{\mu_{f^{y_{k',l'}} \!\to x_{k',l',p',q'}}}{\eta_{f^{y_{k',l'}} \!\to x_{k',l',p',q'}}}.
\end{align}

\begin{figure*}[!t]
\begin{equation}\label{symbol_mean}
\breve{\bar{x}}_{k,l,p,q} =
\begin{cases}
\sum\limits_{c^s \in \mathcal M} \breve{p}(x_{k,l,p,q} = c^s) c^s  & \mbox{if ${x}_{k,l,p,q}$ is a data symbol,}
\\
c^s &\mbox{if ${x}_{k,l,p,q}$ is a pilot symbol with value $c^s$.}
\end{cases}
\end{equation}
\begin{equation}\label{symbol_Var}
\breve{\text{Var}}\left({x}_{k,l,p,q}\right)=
\begin{cases}
\sum\limits_{c^s\in \mathcal M} \breve{p}(x_{k,l,p,q} = c^s)|c^s|^2 - |\breve{\bar{x}}_{k,l,p,q}|^2  & \mbox{if ${x}_{k,l,p,q}$ is a data symbol,}
\\
0&\mbox{if ${x}_{k,l,p,q}$ is a pilot symbol.}
\end{cases}
\end{equation}
\hrulefill
\end{figure*}

Thus, for a data symbol $x_{k,l,p,q}$, its extrinsic PDF is given by:
$
\breve{p}(x_{k,l,p,q} = c^s)
\propto p(x_{k,l,p,q} = c^s)
\mathcal {CN} \left(c^s; \mu_{f^{\mathbf y \setminus y_{k,l}} \to x_{k,l,p,q}}^{sum}, \eta_{f^{\mathbf y \setminus y_{k,l}} \to x_{k,l,p,q}}^{sum} \right)
$.
We further project it into a Gaussian distribution with the mean and variance given by \eqref{symbol_mean} and \eqref{symbol_Var} on the top of this page.
Therefore, by adopting a damping factor $\rho \in (0,1]$, $\mu_{x_{k,l,p,q} \to f^{y_{k,l}}}$ and $\eta_{x_{k,l,p,q} \to f^{y_{k,l}}}$ can be determined with $\{\breve{\bar{x}}_{k,l,p,q}, \breve{\text{Var}}\left({x}_{k,l,p,q}\right)\}$ of current and one previous iterations.

After each iteration within MP, the messages to the variable node ${x}_{k,l,p,q}$ from its connected function nodes $f^{y_{k,l}}$ are updated.
By combining all the messages to the data symbol $x_{k,l,p,q}$, we can obtain the a posterior PDF as
$
\widehat{p}(x_{k,l,p,q} \!\!=\!\! c^s)
\!\propto\!\!\!
\prod\limits_{(k',l'p',q') \in \mathcal J_{y^{k,l}}^{x^{p,q}} }
\!\!\!\!\mathcal {CN} \big(c^s; \!\mu_{f^{y_{k',l'}} \!\to x_{k',l',p',q'}}, \! \eta_{f^{y_{k',l'}} \!\to x_{k',l',p',q'}} \big)
$.
Then the determination of the data symbol can be implemented by selecting the maximum $\widehat{p}(x_{k,l,p,q} = c^s)$ within all $c^s\in \mathcal M$ for each $x_{k,l,p,q}$.

\subsubsection{\bf Left part for equivalent channel gain estimation}

To implement the MP in the left side of the whole factor graph, we collect the receiving grids related to pilot symbols as $\{Y^{\text{DD}}_{k,l} | (k,l) \in \mathcal D_P\}$, where it is also affected by transmitted symbols $\{X^{\text{DD}}_{k,l} | (k,l) \in \mathcal C_{\mathcal D_{P}}\}$.
During BP procedure within the left part of the factor graph,
we only resort to the message from transmitted symbols $\{X^{\text{DD}}_{k,l} | (k,l) \in \mathcal C_{\mathcal D_{P}}\}$ to $\{Y^{\text{DD}}_{k,l}| (k,l) \in \mathcal D_P \}$, for the channel gain estimation.
Similar to the data detection process, we first derive the message from function node $f^{y_{k,l}}$ to variable node $\widetilde{h}_p$ as
\begin{align}
&\vartheta_{ f^{y_{k,l}} \to \widetilde{h}_p}
\notag \\
=& \!\!\!\! \prod_{(p'\!,q'\!)\in \mathcal G_{x}^{y^{k,l}} }
\!\! \sum_{x_{k,l,p'\!,q'} \!\in\! \mathcal M}
\!\!\!\!\!\!\!\!\mathcal {CN}\! \big(x_{k,l,p',q'};\! \mu_{x_{k,l,p'\!,q'} \!\to\! f^{y_{k,l}}}, \!\eta_{x_{k,l,p'\!,q'} \!\to\! f^{y_{k,l}}} \!\big) \notag \\
&\times
\int_{\bar{y}_{k,l}} \mathcal{CN}\left(y_{k,l}; \bar{y}_{k,l}, \sigma_n^2 \right)
\mathcal{CN} \big(\bar{y}_{k,l}; \widehat{\bar{y}}_{k,l}^{\widetilde{h}}, \lambda^{\bar{y},\widetilde{h}}_{k,l} \big)
\notag \\
\propto&
\mathcal {CN} \big(\widetilde{h}_p; \mu_{f^{y_{k,l}} \to{\widetilde{h}_{p}}}, \eta_{f^{y_{k,l}} \to {\widetilde{h}_{p}} } \big),
\end{align}
where the mean and variance of $\bar{y}_{k,l}$ are replaced with
$\widehat{\bar{y}}_{k,l}^{\widetilde{h}} =  \boldsymbol \gamma_{k,l,p}^T \mathbf x_{k,l,p} \widetilde{h}_p + \widehat{u}_{p}^{y,\widetilde{h}}$ and
$\lambda^{\bar{y},\widetilde{h}}_{k,l} = v_{k,l,p}^{u_{\bar{y},\widetilde{h}}}$.
By defining
$\mu_{k,l,p'}^{\gamma x} = \sum\limits_{q'=0}^{N-1} \gamma_{k,l,p',q'} \breve{\bar{x}}_{k,l,p',q'}$ and
$\eta_{k,l,p'}^{\gamma x} = \sum\limits_{q'=0}^{N-1} |\gamma_{k,l,p',q'}|^2 \breve{\text{Var}}\left({x}_{k,l,p',q'}\right)$,
we can obtain
\begin{align}
\widehat{u}_{p}^{y,\widetilde{h}} \!\triangleq&\! \sum_{p' \ne p}
\mu_{k,l,p'}^{\gamma x} \mu_{\widetilde{h}_{p'} \to f^{y_{k,l}}},
\\
v_{k,l,p}^{u_{y,\widetilde{h}}} \!\triangleq&\! \sum\limits_{p' \ne p}
\big(\eta_{{\widetilde{h}_{p'}} \to f^{y_{k,l}}}
\left(\eta_{k,l,p'}^{\gamma x} \!+ |\mu_{k,l,p'}^{\gamma x}|^2 \right)
\notag \\
&+ |\mu_{{\widetilde{h}_{p'}} \to f^{y_{k,l}}}|^2 \eta_{k,l,p'}^{\gamma x}
\big),
\label{part_var_z}
\end{align}
and $\mu_{\widetilde{h}_{p'} \to f^{y_{k,l}}}$ and $\eta_{\widetilde{h}_{p'} \to f^{y_{k,l}}}$ are the mean and variance of the message $\vartheta_{\widetilde{h}_{p'} \to f^{y_{k,l}}}$ from variable node $\widetilde{h}_{p'}$ to function node $f^{y_{k,l}}$ in the last iteration, which will be defined later.
Thus,
$\mu_{f^{y_{k,l}} \to{\widetilde{h}_{p}}} = \frac{y_{k,l} - \widehat{u}_{p}^{y,\widetilde{h}}}
{\mu_{k,l,p}^{\gamma x}}
$,
$
\eta_{f^{y_{k,l}} \to{\widetilde{h}_{p}}} = \frac{v_{k,l,p}^{u_{y,\widetilde{h}}}
+ \sigma_n^2
- |\mu_{f^{y_{k,l}} \to{\widetilde{h}_{p}}}|^2 \eta_{k,l,p}^{\gamma x}
}
{\eta_{k,l,p}^{\gamma x} + |\mu_{k,l,p}^{\gamma x}|^2  }
$.

Combining all the messages from function nodes $f^{y_{k',l'}}$ to variable node $\widetilde{h}_p$ where $(k',l') \in \overline{\mathcal D}_{P}^{k,l} =  \{\mathcal D_P \setminus (k,l)\}$, we can obtain a Gaussian message with the mean and the variance given by $\eta_{f^{\mathbf y/ y_{k,l}} \to{\widetilde{h}_{p}}}^{sum}
= \big(\sum\limits_{(k',l') \in \overline{\mathcal D}_{P}^{k,l}} \frac{1}{\eta_{f^{y_{k',l'}} \to {\widetilde{h}_{p}} }}\big)^{-1}$ and $\mu_{f^{\mathbf y/ y_{k,l}} \to{\widetilde{h}_{p}}}^{sum} =
\eta_{f^{\mathbf y/ y_{k,l}} \to{\widetilde{h}_{p}}}^{sum}
\sum\limits_{(k',l') \in \overline{\mathcal D}_{P}^{k,l} } \frac{\mu_{f^{y_{k',l'}} \to{\widetilde{h}_{p}}}}{\eta_{f^{y_{k',l'}} \to{\widetilde{h}_{p}}}}$, respectively.
So the message from variable node $\widetilde{h}_{p} $ to function node $f^{y_{k,l}}$ can be expressed as
$
\vartheta_{ \widetilde{h}_p \to f^{y_{k,l}}}
= p(\widetilde{h}_{p})\prod\limits_{(k',l') \in \overline{\mathcal D}_{P}^{k,l} } \vartheta_{ f^{y_{k',l'}} \to \widetilde{h}_p}
\approx
\mathcal{CN} \big(\widetilde{h}_{p}; \mu_{ \widetilde{h}_p \to f^{y_{k,l}}}, \eta_{ \widetilde{h}_p \to f^{y_{k,l}}} \big)
$,
where
$
\mu_{ \widetilde{h}_p \to f^{y_{k,l}}}
=
\frac{\lambda_{p} \mu_{f^{\mathbf y/ y_{k,l}} \to{\widetilde{h}_{p}}}^{sum}}
{\lambda_{p} + \eta_{f^{\mathbf y/ y_{k,l}} \to{\widetilde{h}_{p}}}^{sum}} K_{k,l,p}
$,
$
\eta_{ \widetilde{h}_p \to f^{y_{k,l}}}
=
\frac{\lambda_{p} \eta_{f^{\mathbf y/ y_{k,l}} \to{\widetilde{h}_{p}}}^{sum}}
{\lambda_{p} + \eta_{f^{\mathbf y/ y_{k,l}} \to{\widetilde{h}_{p}}}^{sum}} K_{k,l,p} + \Big|\frac{\lambda_{p} \mu_{f^{\mathbf y/ y_{k,l}} \to{\widetilde{h}_{p}}}^{sum}}
{\lambda_{p} + \eta_{f^{\mathbf y/ y_{k,l}} \to{\widetilde{h}_{p}}}^{sum}} \Big|^2 (K_{k,l,p}-K_{k,l,p}^2)
$,
and $K_{k,l,p} \triangleq \frac{\alpha \mathcal{CN} \big(0; \mu_{f^{\mathbf y/ y_{k,l}} \to{\widetilde{h}_{p}}}^{sum},  \lambda_{p} + \eta_{f^{\mathbf y/ y_{k,l}} \to{\widetilde{h}_{p}}}^{sum}\big)}{\int_{\widetilde{h}_p} p(\widetilde{h}_{p})\prod\limits_{(k',l') \in \overline{\mathcal D}_{P}^{k,l}} \vartheta_{ f^{y_{k',l'}} \to \widetilde{h}_p}}$.

After each iteration of MP, the messages to the variable node $\widetilde{h}_p$ from function nodes $f^{\mathbf y}$  are updated.
Then the a posterior PDF of the equivalent channel gains can be derived as:
$
\breve{p}(\widetilde{h}_p) \propto {p}(\widetilde{h}_p)
\prod\limits_{(k,l) \in \mathcal D_P} \vartheta_{ f^{y_{k,l}} \to \widetilde{h}_p}
= \mathcal {CN} \big(\widetilde{h}_p; \widehat{\widetilde{h}}_p,  {\breve{\lambda}}_p \big)
$,
where ${\breve{\lambda}}_p$ is the a posterior variance of ${\widetilde{h}}_p$.
Define $\eta_{f^{\mathbf y} \to{\widetilde{h}_{p}}}^{sum}
= \Big(\sum\limits_{(k,l) \in \mathcal D_P} \frac{1}{\eta_{f^{y_{k,l}} \to {\widetilde{h}_{p}} }}\Big)^{-1}$ and $\mu_{f^{\mathbf y} \to{\widetilde{h}_{p}}}^{sum} =
\eta_{f^{\mathbf y} \to{\widetilde{h}_{p}}}^{sum}
\sum\limits_{(k,l) \in \mathcal D_P} \frac{\mu_{f^{y_{k,l}} \to{\widetilde{h}_{p}}}}{\eta_{f^{y_{k,l}} \to{\widetilde{h}_{p}}}}$,
then we can obtain
\begin{align}
\widehat{\widetilde{h}}_p =&
\frac{\lambda_{p} \mu_{f^{\mathbf y} \to{\widetilde{h}_{p}}}^{sum}}
{\lambda_{p} + \eta_{f^{\mathbf y} \to{\widetilde{h}_{p}}}^{sum}} K_{p},
\\
{\breve{\lambda}}_p
=&
\frac{\lambda_{p} \eta_{f^{\mathbf y} \to{\widetilde{h}_{p}}}^{sum}}
{\lambda_{p} + \eta_{f^{\mathbf y} \to{\widetilde{h}_{p}}}^{sum}} K_{p} + \Big|\frac{\lambda_{p} \mu_{f^{\mathbf y} \to{\widetilde{h}_{p}}}^{sum}}
{\lambda_{p} + \eta_{f^{\mathbf y} \to{\widetilde{h}_{p}}}^{sum}} \Big|^2 (K_{p}-K_{p}^2),
\end{align}
where $K_{p} \triangleq \frac{\alpha \mathcal{CN} \big(0; \mu_{f^{\mathbf y} \to{\widetilde{h}_{p}}}^{sum},  \lambda_{p} + \eta_{f^{\mathbf y} \to{\widetilde{h}_{p}}}^{sum} \big)}{\int_{\widetilde{h}_p} p(\widetilde{h}_{p})\prod\limits_{(k,l) \in \mathcal D_P} \vartheta_{ f^{y_{k,l}} \to \widetilde{h}_p}}$.

\begin{remark}
In the MP procedure,
since we only take the possible received grids of pilot symbols into consideration for the equivalent channel gain estimation,
only the messages from the data symbols located in the interference area are utilized.
This can enhance the channel estimation accuracy in the initial iteration.
With better initial value, the proposed JCEDD scheme will converge faster, resulting in a low complexity.
\end{remark}

\subsection{EM based Parameters Updating}

\subsubsection{\bf E-step}
The EM algorithm is divided into two steps, i.e., the expectation step (E-step) and the maximization step (M-step).
Within the E-step, the objective function is derived as
\begin{align}\label{E_function}
Q(\boldsymbol \Xi, \widehat{\boldsymbol \Xi}^{(l-1)})
=& \mathbb E_{\widetilde{\mathbf h}|\mathbf y, \widehat{\boldsymbol \Xi}^{(l-1)}}
\left\{\ln p\left(\mathbf y \Big| \mathbf Z, \widetilde{\mathbf h}\right)
\right\}
\notag \\
&+ \mathbb E_{\widetilde{\mathbf h}|\mathbf y, \widehat{\boldsymbol \Xi}^{(l-1)}}
\left\{\ln p\left(\widetilde{\mathbf h} \right)\right\}.
\end{align}

It can be checked that the Doppler-related parameters $\{k_{\nu_p}, \beta_{\nu_p}\}_{p=1}^{P_{max}}$ only depend on the first item of \eqref{E_function}, while the characteristic of channel gain, i.e. $\alpha$ and $\mathbf \Lambda$, only affect the second item.
By further derivations, we can represent the expectation functions respectively for $\{k_{\nu_p}, \beta_{\nu_p}\}$ and $\{\alpha, \mathbf \Lambda\}$ in the $t$-th EM iteration as:
\begin{align}
&Q(k_{\nu_p}, \beta_{\nu_p}, \widehat{\boldsymbol \Xi}^{(t-1)})
\notag \\
=& \sum_{(k,l) = (-\frac{N}{2},0)}^{(\frac{N}{2}-1, M-1)}
2\Re\left\{ ( y_{k,l})^*
\mathbb E \big\{\widetilde{h}_p|\mathbf y, \widehat{\boldsymbol \Xi}^{(t-1)} \big\}
\boldsymbol{\gamma}_{k,l,p}^T \mathbf x_{k,l,p} \right\} \notag \\
&-
\sum_{(k,l) = (-\frac{N}{2},0)}^{(\frac{N}{2}-1, M-1)}
2\Re\bigg\{ \sum_{p' \neq p}
\left(\boldsymbol{\gamma}_{k,l,p}^T \mathbf x_{k,l,p} \right)^{*}
(\boldsymbol{\gamma}_{k,l,p'}^T \mathbf x_{k,l,p'})
\notag \\
&\times
\mathbb E \Big\{\widetilde{h}_p^* |\mathbf y, \widehat{\boldsymbol \Xi}^{(t-1)} \Big\}
 \mathbb E \big\{\widetilde{h}_{p'}|\mathbf y, \widehat{\boldsymbol \Xi}^{(t-1)} \big\}
\bigg\} \notag \\
&- \!\!\sum_{(k,l) = (-\frac{N}{2},0)}^{(\frac{N}{2}-1, M-1)}\!\!
|\boldsymbol{\gamma}_{k,l,p}^T \mathbf x_{k,l,p}|^2
\mathbb E\left\{\widetilde{h}_p \widetilde{h}^{*}_p| \mathbf y, \widehat{\boldsymbol \Xi}^{(t-1)} \right\} \!\!+\!\!C,
\label{E_function_tau_nu}
\\
&Q(\alpha, \mathbf \Lambda, \widehat{\boldsymbol \Xi}^{(t-1)})
\notag \\
=& \sum_{p=1}^{P_{max}}\!\!
\mathbb E_{\widetilde{\mathbf h}|\mathbf y, \widehat{\boldsymbol \Xi}^{(t-1)}}
\ln \left(\! (1 \!- \alpha)\delta(\widetilde h_{p}) \!+\! \alpha \mathcal{CN}(\widetilde h_{p}; 0, \lambda_{p})\right),
 \label{E_function_alpha_lambda}
\end{align}
where $C$ is the terms not related with $k_{\nu_p}$ and $\beta_{\nu_p}$.
Notice that we omit the top mark of the estimated $\widehat{\mathbf x}_{k,l,p}$ and $\widehat{\widetilde{\mathbf h}}$ for notation simplicity.
Besides, from the result of MP procedure, we have $\mathbb E\left\{\widetilde{h}_p | \mathbf y, \widehat{\boldsymbol \Xi} \right\} = \widehat{\widetilde{h}}_p$ and $\mathbb E\left\{\widetilde{h}_p \widetilde{h}^{*}_p| \mathbf y, \widehat{\boldsymbol \Xi} \right\} = |\widehat{\widetilde{h}}_p|^2 + {\breve{\lambda}}_p$.
In the M-step, we will maximize $Q(\boldsymbol \Xi, \widehat{\boldsymbol \Xi}^{(t-1)})$ by deriving the optimal parameters in $\boldsymbol \Xi$ one by one.

\subsubsection{\bf M-step for updating $k_{\nu_p}$ and $\beta_{\nu_p}$}
For $k_{\nu_p}$, it is obvious that it should be chosen within the set $\mathbf \Omega_{k_{\nu}} = \{-N_{\nu_{max}}, \ldots, 0, \ldots, N_{\nu_{max}} \}$, where $k_{\nu_{max}}$ is the integer grid with the maximum possible Doppler frequency shift.
Thus, the estimation of $k_{\nu_p}$ can be derived by searching the grids in $\mathbf \Omega_{k_{\nu}}$ as
$
\widehat{k}_{\nu_p}^{(t)} = \arg \max_{k_{\nu_p} \in \mathbf \Omega_{k_{\nu}}}
Q(k_{\nu_p}, \widehat{\beta}_{\nu_p}^{(t-1)})
$.

For updating $\beta_{\nu_p}$,
we calculate the derivative of $Q({\beta}_{\nu_p}, \widehat{k}_{\nu_p}^{(t-1)})$ with respect to $\beta_{\nu_p}$ as equation \eqref{deriv_beta} in Appendix A before proceeding.
However, it is difficult to find a closed-form solution for ${\beta}_{\nu_p}$.
Instead, we resort to the gradient ascent algorithm to search the sub-optimal $\widehat{\beta}_{\nu_p}$.
Then, define an initial search point $\beta_{\nu_p}^{(t)(0)}$ and an initial step length $\kappa^{(1)}$,
the initial searching direction can be denoted as $\omega_1 = \frac{\partial Q(\beta_{\nu_p}, \widehat{k}_{\nu_p}^{(t-1)})}{\partial \beta_{\nu_p}} |_{\beta_{\nu_p} = \beta_{\nu_p}^{(t)(0)}}$.
Subsequently, the following searching points can be updated by $\beta_{\nu_p}^{(t)(m+1)} = \beta_{\nu_p}^{(t)(m)} + \kappa^{(m+1)} \omega_m$.
During the $m$-th searching process, if $Q(\beta_{\nu_p}^{(t)(m)}, \widehat{k}_{\nu_p}^{(t-1)}) \leq Q(\beta_{\nu_p}^{(t)(m-1)}, \widehat{k}_{\nu_p}^{(t-1)})$,
then the step length should be updated as $\kappa^{(m)} = \xi\kappa^{(m-1)}$ for better performance, where $\xi$ is an attenuation factor and can be chosen within the range $(0,1)$.
Let $\zeta$ denotes the threshold value.
If $|Q(\beta_{\nu_p}^{(t)(m)}) - Q(\beta_{\nu_p}^{(t)(m-1)})| \leq \zeta$, the algorithm will be terminated.
After the searching process, the estimate $\widehat{\beta}_{\nu_p}^{(t)}$ can be obtained.

\subsubsection{\bf M-step for updating $\alpha$ and $\mathbf \Lambda$}
As for the other two parameters $\alpha$ and $\mathbf \Lambda$, we can update them according to the method in \cite{EMBGAMP}.
With $N_O$ OTFS blocks, the value of $\lambda_p$ and $\alpha$ is necessarily a value that zeroes the corresponding derivative of \eqref{E_function_alpha_lambda}, i.e.,
\begin{align}
&\sum_{n_o = 1}^{N_O}
\int_{\widetilde{h}_p^{(n_o)}} p\left(\widetilde{h}_p | \mathbf y^{(n_o)}; \widehat{\boldsymbol{\Xi}}^{(t-1)} \right)
\notag \\
&\times
\frac{\partial}{\partial \lambda_p}\!\!\ln p \!\left(\widetilde{h}_p^{(n_o)}; \widehat{\boldsymbol{\Xi}}^{(t-1)} \!\!\setminus\! \mathbf \lambda_p \!\right) \!=\! 0, p \!=\! 1,2,\ldots, P_{max},
\label{int_h_p_lambda}\\
&\sum_{n_o = 1}^{N_O}
\sum_{p=1}^{P_{max}} \int_{\widetilde{h}_p^{(n_o)}} p\left(\widetilde{h}_p^{(n_o)} | \mathbf y; \widehat{\boldsymbol{\Xi}}^{(t-1)} \right)
\notag \\
&\times
\frac{\partial}{\partial \alpha}\ln p \left(\widetilde{h}_p^{(n_o)}; \widehat{\boldsymbol{\Xi}}^{(t-1)}\setminus \alpha \right) = 0.
\label{int_h_p_alpha}
\end{align}

The derivative of $p(\widetilde{h}_p^{(n_o)}; \hat{\boldsymbol{\Xi}}^{(t-1)})$ with respect to $\lambda_p $ and $\alpha$ can be further represented as
\begin{align}\label{deriv_lambda}
&\frac{\partial}{\partial\lambda_p}\ln p \left(\widetilde{h}_p^{(n_o)}; \widehat{\boldsymbol{\Xi}}^{(t-1)}\setminus \mathbf \lambda_p^{(t-1)} \right)
\notag \\
&= \frac{1}{2}\left(\frac{|\widetilde{h}_p^{(n_o)}|^2}{\lambda_p^2}- \frac{1}{\lambda_p}\right)
\frac{\alpha^{(t-1)} \mathcal{CN}\left(\widetilde{h}_p^{(n_o)}; 0; \lambda_p  \right)}{p(\widetilde{h}_p^{(n_o)}; \widehat{\boldsymbol{\Xi}}^{(t-1)}\setminus \lambda_p^{(t-1)})} \notag \\
&=
\begin{cases}
\!\frac{1}{2} \!\!\left(\frac{|\widetilde{h}_p^{(n_o)}|^2}{\lambda_p^2}\!-\! \frac{1}{\lambda_p}\!\right) &\widetilde{h}_p^{(n_o)} \!\!\neq\! 0
\\
\!0 &\widetilde{h}_p^{(n_o)} \!\!=\! 0
\end{cases}
, p \!=\! 1,2,\ldots, P_{max},
\end{align}
\begin{align}\label{deriv_alpha}
&\frac{\partial}{\partial \alpha}\ln p \left(\widetilde{h}_p^{(n_o)}; \widehat{\boldsymbol{\Xi}}^{(t-1)}\setminus \alpha^{(t-1)} \right) \notag \\
&= \frac{\mathcal{CN}\left(\widetilde{h}_p^{(n_o)}; 0, \lambda_p^{(t-1)} \right) - \delta(\widetilde{h}_p)}{p \left(\widetilde{h}_p^{(n_o)}; \widehat{\boldsymbol{\Xi}}^{(t-1)} \setminus \alpha^{(t-1)} \right)}
\notag \\
&=
\begin{cases}
\frac{1}{\alpha}
&\widetilde{h}_p^{(n_o)} \neq 0,
\\
\frac{-1}{1-\alpha}
&\widetilde{h}_p^{(n_o)} = 0.
\end{cases}
\end{align}

Hence, the neighborhood around the point $\widetilde{h}_p = 0$ should be treated differently than the remainder of $\mathbb C$.
Thus, we define the closed ball $\mathcal B_\epsilon = \{x| |x-0|\leq \epsilon\}$ and $\overline{\mathcal B}_{\epsilon} \triangleq \mathbb C \setminus \mathcal B_\epsilon$.
By splitting the domain of integration in \eqref{int_h_p_lambda} into $\mathcal B_{\epsilon}$ and $\overline{\mathcal B}_{\epsilon}$, and then plugging in \eqref{deriv_lambda},
we find that the following is equivalent to \eqref{int_h_p_lambda} in the limit of $\epsilon \to 0$:
\begin{align}\label{maxi_lambda_p}
\sum_{n_O=1}^{N_O}\!\!
\int_{\widetilde{h}_{p}^{(n_o)} \in \overline{\mathcal{B}}_{\epsilon}}\!\!
\left(|\widetilde{h}_{p}^{(n_o)}|^{2}\!-\!\lambda_p\right)
p\left(\widetilde{h}_p^{(n_o)} | \mathbf y; \widehat{\boldsymbol{\Xi}}^{(t-1)} \!\right)
\!= \!0.
\end{align}
The unique value of $\lambda_p$ for $p = 1,2,\ldots, P_{max}$ satisfying \eqref{maxi_lambda_p} as $\epsilon \to 0$ is then
\begin{align}\label{maxi_lambda_p2}
\widehat\lambda_p^{(t)}
=& \frac{1}{\alpha^{(t-1)}N_O}
\sum_{n_O = 1}^{N_O}
\pi_{p}(\mu_{f^{\mathbf y^{(n_o)}} \to{\widetilde{h}_{p}^{(n_o)}}}^{sum}, \eta_{f^{\mathbf y^{(n_o)}} \to{\widetilde{h}_{p}^{(n_o)}}}^{sum}; \boldsymbol{\Xi})
\notag\\
&\times
\left(\left|\Gamma_p(\mu_{f^{\mathbf y^{(n_o)}} \to{\widetilde{h}_{p}^{(n_o)}}}^{sum}, \eta_{f^{\mathbf y^{(n_o)}} \to{\widetilde{h}_{p}^{(n_o)}}}^{sum}; \boldsymbol{\Xi})\right|^2 \right. \notag\\
&\left. + \Theta_p(\mu_{f^{\mathbf y^{(n_o)}} \to{\widetilde{h}_{p}^{(n_o)}}}^{sum}, \eta_{f^{\mathbf y^{(n_o)}} \to{\widetilde{h}_{p}^{(n_o)}}}^{sum}; \boldsymbol{\Xi})\right),
\end{align}
where
$\Gamma_p(\mu_{f^{\mathbf y} \to{\widetilde{h}_{p}}}^{sum}, \eta_{f^{\mathbf y} \to{\widetilde{h}_{p}}}^{sum}; \boldsymbol{\Xi})
=
\frac{\widehat\lambda_{p}^{(t-1)} \mu_{f^{\mathbf y} \to{\widetilde{h}_{p}}}^{sum}}
{\widehat\lambda_{p}^{(t-1)} + \eta_{f^{\mathbf y} \to{\widetilde{h}_{p}}}^{sum}}$,
$\Theta_p(\mu_{f^{\mathbf y} \to{\widetilde{h}_{p}}}^{sum}, \eta_{f^{\mathbf y} \to{\widetilde{h}_{p}}}^{sum}; \boldsymbol{\Xi})
=
\frac{\widehat\lambda_{p}^{(t-1)} \eta_{f^{\mathbf y} \to{\widetilde{h}_{p}}}^{sum}}
{\widehat\lambda_{p}^{(t-1)} + \eta_{f^{\mathbf y} \to{\widetilde{h}_{p}}}^{sum}}$,
and
\begin{align}\label{pi_p}
&\pi_{p}(\mu_{f^{\mathbf y} \to{\widetilde{h}_{p}}}^{sum}, \eta_{f^{\mathbf y} \to{\widetilde{h}_{p}}}^{sum}; \boldsymbol{\Xi})
\notag\\
&\triangleq
\frac{\widehat\alpha^{(t-1)} \mathcal{CN} \left(0; \mu_{f^{\mathbf y} \to{\widetilde{h}_{p}}}^{sum},  \widehat\lambda_{p}^{(t-1)} + \eta_{f^{\mathbf y} \to{\widetilde{h}_{p}}}^{sum}\right)}
{\int_{\widetilde{h}_p} p(\widetilde{h}_{p})\prod\limits_{(k,l)} \vartheta_{ f^{y_{k,l}} \to \widetilde{h}_p}}.
\end{align}


Similarly, Plugging \eqref{deriv_alpha} into \eqref{int_h_p_alpha} with the limit $\epsilon \to 0$, we can obtain
\begin{align}
&\frac{1}{\alpha}\!\! \sum_{n_O = 1}^{N_O}\!\!\! \sum_{p=1}^{P_{max}}
\!\!\lim\limits_{\epsilon \to 0} \int_{\widetilde{h}_{p}^{(n_o)} \!\in \overline{\mathcal{B}}_{\epsilon}}
\!\!\!p \!\left(\widetilde{h}_p^{(n_o)} | \mathbf y^{(n_o)}; \!\hat{\boldsymbol{\Xi}}^{(t-1)} \!\right)
\notag \\
&=\!\!
\frac{1}{1\!-\!\alpha} \!\!\sum_{n_O = 1}^{N_O}\!\!\! \sum_{p=1}^{P_{max}} \!\lim\limits_{\epsilon \to 0} \!\!\int_{\widetilde{h}_{p}^{(n_o)} \in \mathcal{B}_{\epsilon}}
\!\!\!p \!\left(\widetilde{h}_p^{(n_o)} | \mathbf y^{(n_o)}; \!\hat{\boldsymbol{\Xi}}^{(t-1)} \!\right),
\end{align}
and finally $\widehat\alpha^{(t)}$ can be updated by
$
\widehat\alpha^{(t)} = \frac{1}{P_{max}N_O}
\sum_{p=1}^{P_{max}}
\sum_{n_O = 1}^{N_O}
\pi_{p}(\mu_{f^{\mathbf y} \to{\widetilde{h}_{p}^{(n_o)}}}^{sum}, \eta_{f^{\mathbf y^{(n_o)}} \to{\widetilde{h}_{p}^{(n_o)}}}^{sum}; \boldsymbol{\Xi})
$.

By employing the above updated $\widehat{\boldsymbol \Xi}$ as known parameters, we can feed them back to the MP module, and form an iterative procedure.
After a certain number of iterations,
both the CSI and the data symbols can be simultaneously acquired.
\subsection{Complexity Analysis}
Consider the OTFS transmission pattern of \figurename{ \ref{old_pattern1}} in the manuscript.
Thus, there are $M_PN_P$ pilot symbols and $(M-M_P)N$ data symbols within one OTFS block.
Then, it is obvious that the scale of observation area for the pilot and data are respectively $(M_P+\frac{\tau_{max}}{T_s})N$ and $(M-M_P+\frac{\tau_{max}}{T_s})N$.
With the system configuration in the simulation result, the observation area for the data can be $MN$.
Assume the number of JCEDD iterations is $N_{iter}$.
For each JCEDD iteration, the computational complexity is divided into two parts, i.e., the EM and the MP.
Note that the EM process only contains scalar addition and multiplication of constant times, and its computation complexity can be easily derived as $\mathcal O(MN^2)$.
Then, focusing on the MP process of one OTFS block, the main computation complexity comes from the calculation of the messages.
The computational complexity of the message from the observation function nodes to all the factor nodes of data symbols is $\mathcal O(MN^2P_{max})$.
The computational complexity of the message from factor nodes of all the data symbols to the observation function nodes is $\mathcal O(|\mathcal M| (M-M_P)(P_{max}N-1)MN^3P_{max})$.
Since $M_P$ is very small compared to $M$ in the symbol pattern of our proposed scheme, the total computational complexity of the data detection part for each iteration is $\mathcal O(|\mathcal M|P_{max}^2M^2N^4)$.
For the equivalent channel gain estimation part, the computational complexity is $\mathcal O(((M_P+\frac{\tau_{max}}{T_s})N-1)P_{max}(M_P+\frac{\tau_{max}}{T_s})N + |\mathcal M|NP_{max}(P_{max}-1)((M_P+\frac{\tau_{max}}{T_s})N-1) )$.
Since we have $P_{max} = \tau_{max} /T_s + 1 $ in the first paragraph of Section IV, the above computational complexity of the equivalent channel gain estimation part for each iteration can be simplified as $\mathcal O((M_P+P_{max})^2 N^2 P_{max} + |\mathcal M|N^2P_{max}^2((M_P+P_{max})) )$.
Finally, the overall computational complexity is $\mathcal O(N_{iter} N_O |\mathcal M|P_{max}^2M^2N^4)$.

\section{Simulation Results}
\subsection{System Settings}
\begin{figure}[htbp]
  \centering
  \subfigure[]
  {\label{old_pattern1}
  \begin{minipage}{70mm}
  \centering
  \includegraphics[width=70mm]{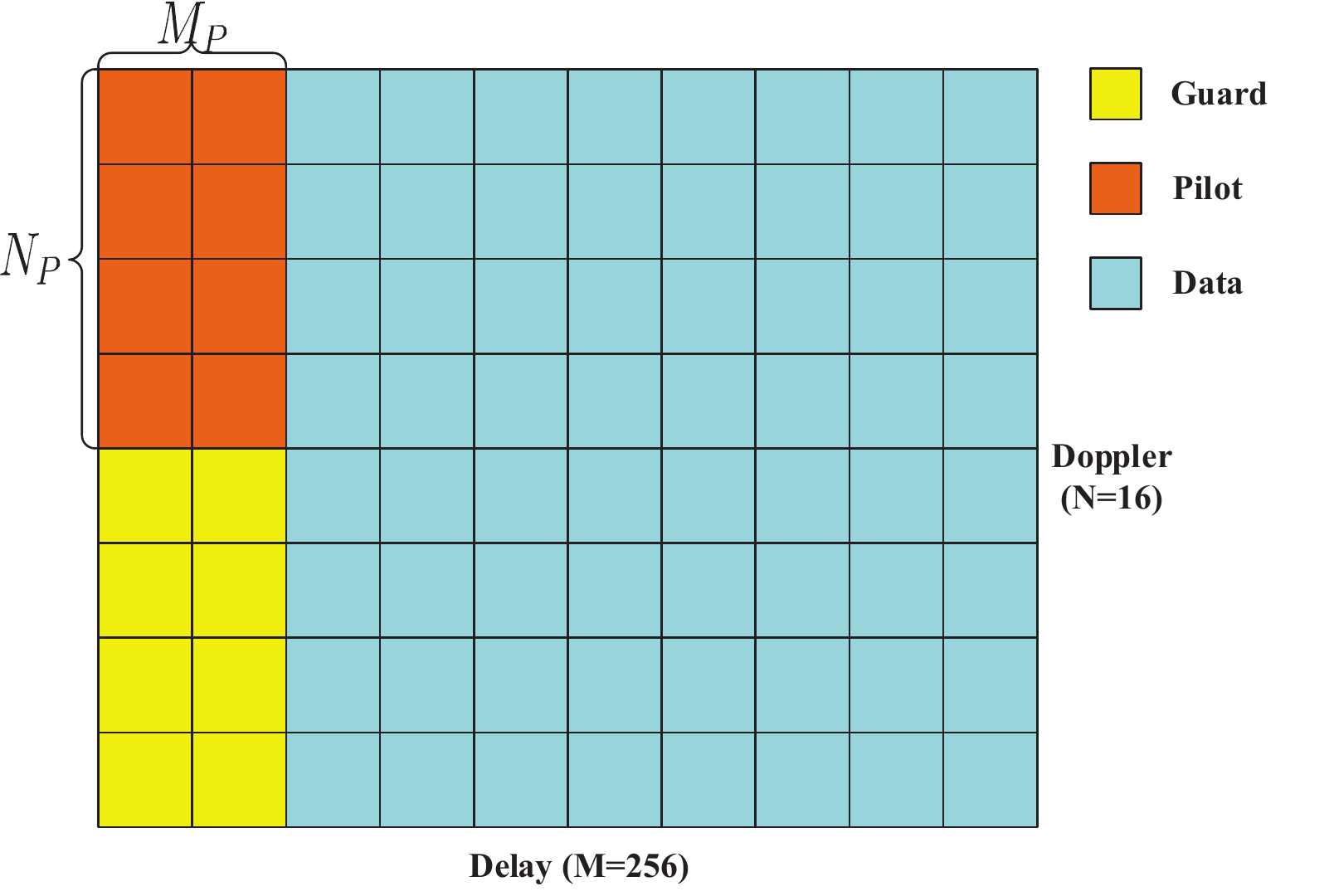}
  \end{minipage}
  }
  \subfigure[]
  {\label{old_pattern2}
  \begin{minipage}{70mm}
  \centering
  \includegraphics[width=70mm]{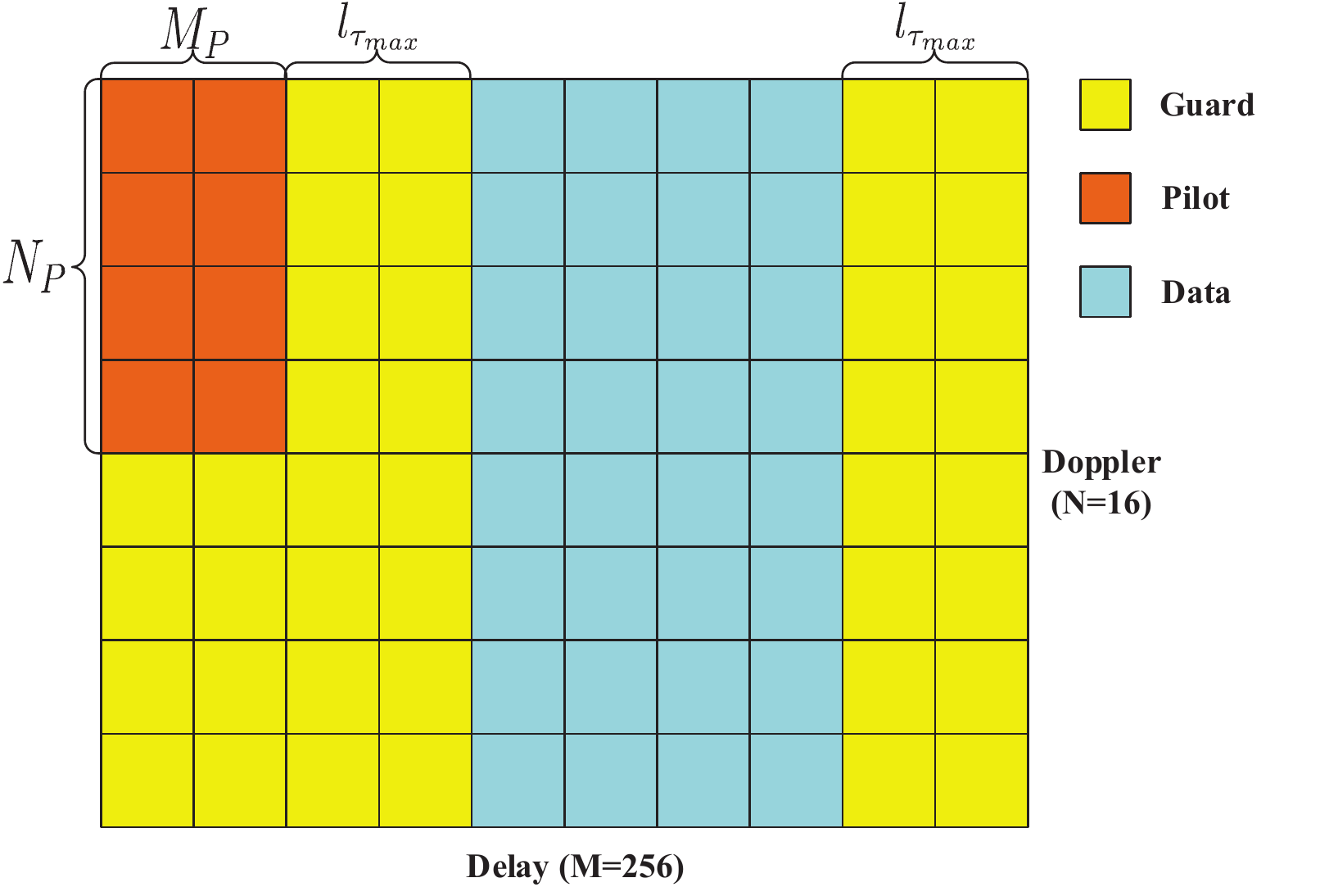}
  \end{minipage}
  }
  \caption{Symbol pattern over delay-Doppler domain: (a) proposed pattern; (b) pattern with sufficient guard.}
  \label{old_pattern}
\end{figure}

We consider an HRIS aided mmWave communication system where the BS is equipped with $N_b = 8$ antennas.
Also, the number of HRIS elements is $N_r = N_x\times N_y = 64\times64$.
$N_{RF} = 8$ is the number of RF chains equipped on the HRIS.
The carrier frequency is $28$ GHz, and the antenna/element spacing $d$ is set as half the wavelength.
$T_s = \frac{1}{20 \text{ MHz}}$ is the system sampling time period.
The user moves with the velocity $v_U \in [0, 240]$ km/h, and thus the maximal Doppler frequency shift is about $\nu_{max} = 6.222$ kHz.
The number of scattering paths from the user to the HRIS is $P=3$.
On the other hand, since only the LoS link is considered from the HRIS to the BS, we assume that
$\tau^{\text{RB}} = 0$.
Thus we have $\tau_p = \tau_p^{\text{UR}}$.
Moreover, the exponentially decaying power delay profile $\sigma_{h^{\text{UR}}_p}^2 = \sigma_c^2 e^{-{\tau_p^{\text{UR}}} / {\tau_{max}^{\text{UR}}}}$ is adopted for $h^{\text{UR}}_p$, where the constant $\sigma_c^2$ is chosen such that the average cascaded channel power is normalized to unity.
The maximum delay spread is set as $\tau_{max} = 6 T_s$.
With respect to the OTFS modulation, the number of the subcarriers is $M = 256$, and the number of time slots is $N = 16$.
Correspondingly, the resolutions over the delay and the Doppler domains are $T_s$ and $\frac{1}{MNT_s} \approx 4.88$ kHz, respectively.
The range of the pilot area is set as $N_P = N/2$ and $M_P = \tau_{max}/T_s$.
The symbol pattern is presented in \figurename{ \ref{old_pattern1}}.
Note that the guard grids within Doppler domain is to
keep the adaptability of the pattern in extremely high mobility scenario,
and the multiple columns of pilot is to enlarge the observation area with sufficient diversity for channel estimation.
Besides, the power gap factor between the pilot and data is set as $\Delta = 0,3,6$ dB.

For the preamble, the length of pilot sequence is $N_t = 16$.
Besides, the number of pilot blocks is $N_B = 2N_x / N_{RF} = 16$.
Hence, the total duration of the preamble is $N_B N_t T_s = 256 T_s$, which is much shorter than that of one OTFS block.
As has been analyzed in Section III, we will adopt one preamble for $N_O = 20$ OTFS blocks.
The SNR is expressed as $\text{SNR} = 10\log_{10} \sigma_r^2/\sigma_n^2$, where $\sigma_r^2$ is the average power for the effective signal at the grids of $N_O$ OTFS blocks over the delay-Doppler domain.
Here, we use the normalized mean square error (NMSE) for the estimated parameters at both the HRIS and the BS, which is defined as
$
\text{MMSE}_{\mathbf x} = \mathbb E
\left\{ \frac{\|\widehat{\mathbf x} - \mathbf x\|^2}{\|\mathbf x\|^2}
 \right\}, \mathbf x = \boldsymbol \tau, \boldsymbol \nu, \boldsymbol \beta, \widetilde{\mathbf h}
$,
with $\widehat{\mathbf x}$ as the estimate of $\mathbf x$.
Besides, the data detection performance is evaluated by the bit error rate (BER).

\subsection{Numerical Results}

\begin{figure}[htbp]
	\centering
	\includegraphics[width=70mm]{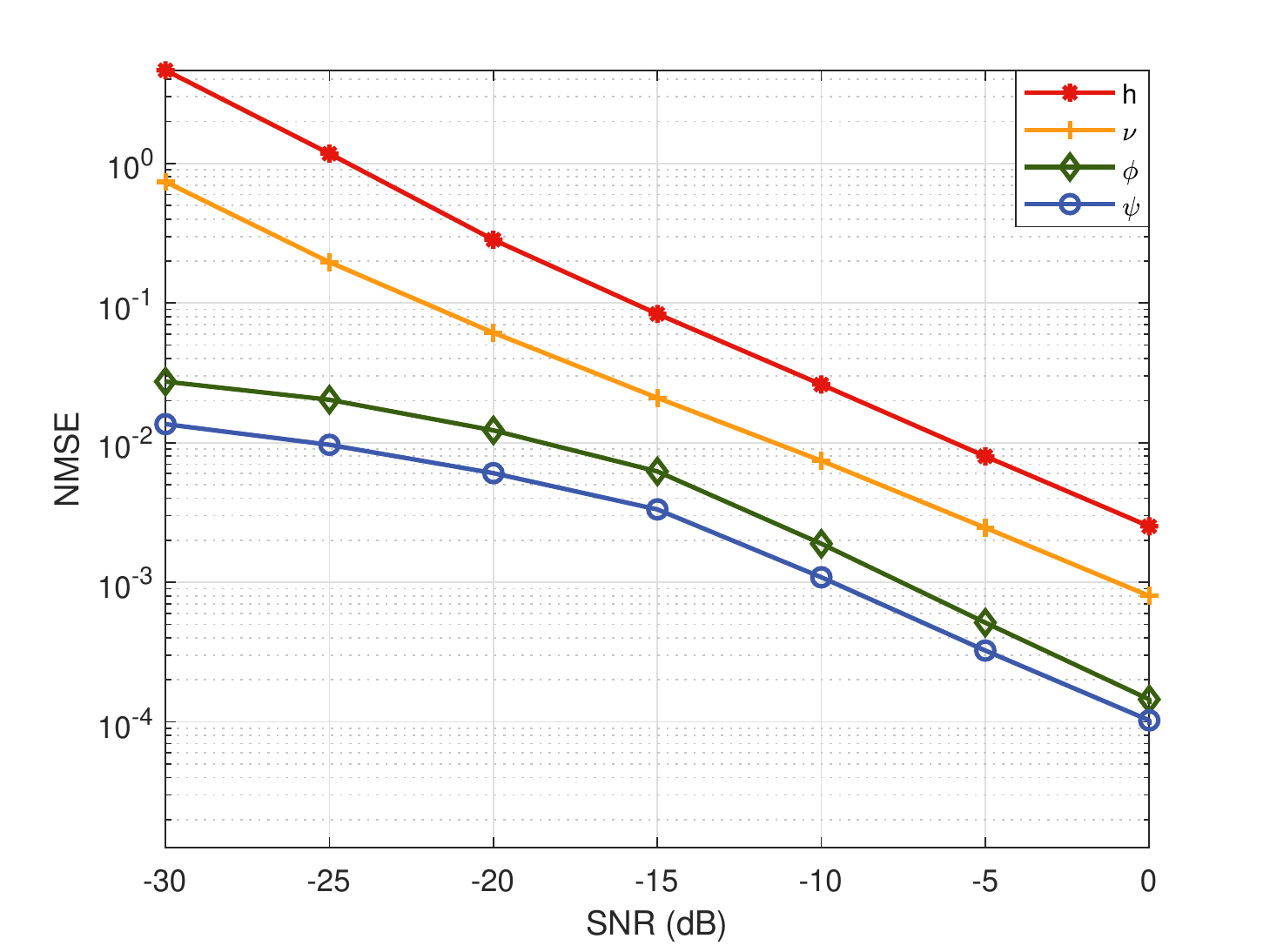}
	\caption{Performance of NOMP based parameter learning at the HRIS.}
	\label{NOMP_MSE_paras_vs_SNR}
\end{figure}

Firstly, we examine the performance of NOMP based parameter learning on the HRIS.
\figurename{ \ref{NOMP_MSE_paras_vs_SNR}} shows the NMSE of the channel parameters versus SNR.
It can be observed that the NMSEs decrease with the increase of SNR linearly, and the parameters are very precise at $\text{SNR} = 0$ dB.
This can help design more efficient beamforming matrix.

\begin{figure}[htbp]
  \centering
  \subfigure[]
  {\label{QPSKMSE_fulldata_SNR15_20_vs_iter}
  \begin{minipage}{70mm}
  \centering
  \includegraphics[width=70mm]{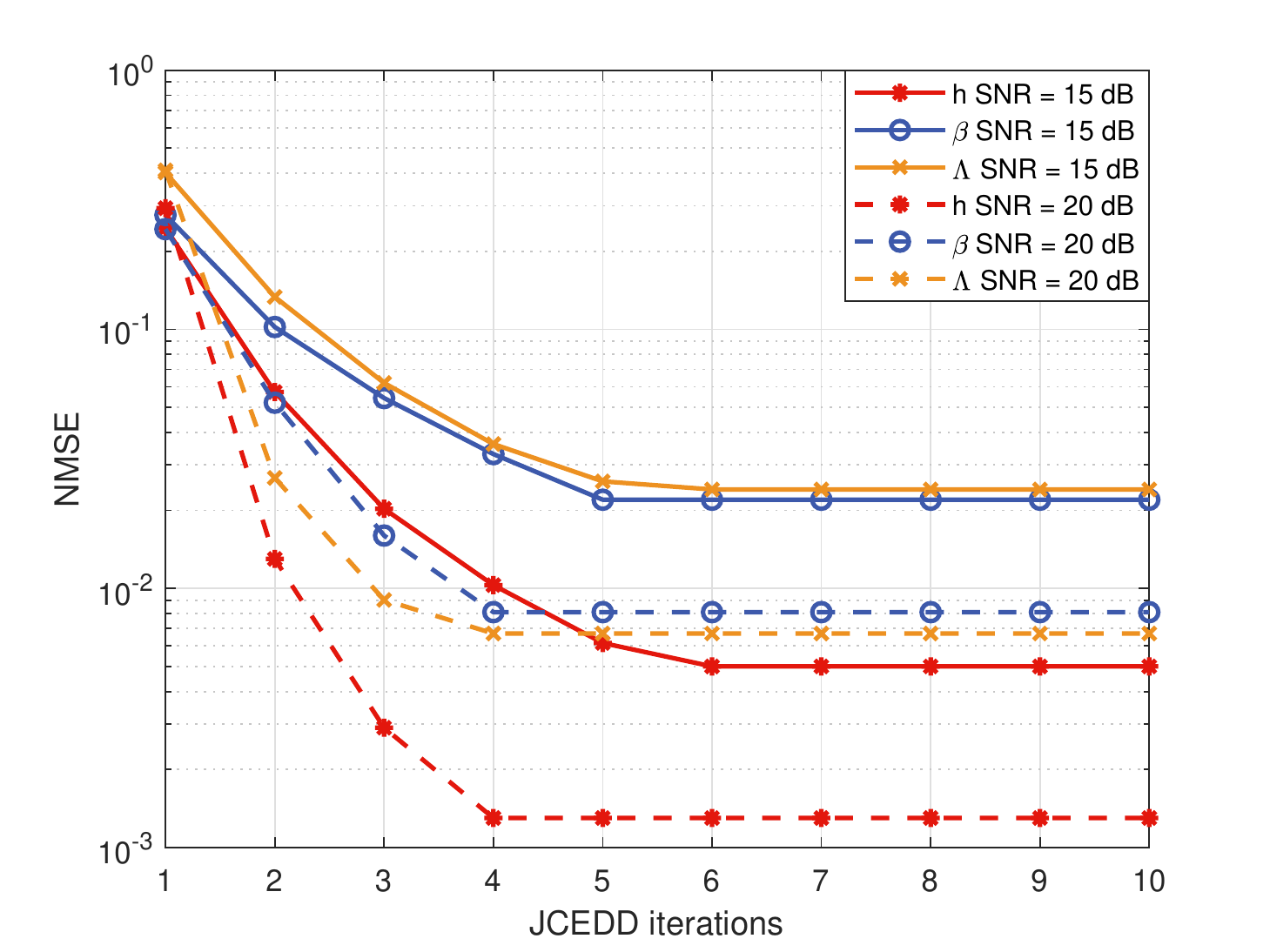}
  \end{minipage}
  }
  \subfigure[]
  {\label{QPSKBER_fulldata_SNR15_20_vs_iter}
  \begin{minipage}{70mm}
  \centering
  \includegraphics[width=70mm]{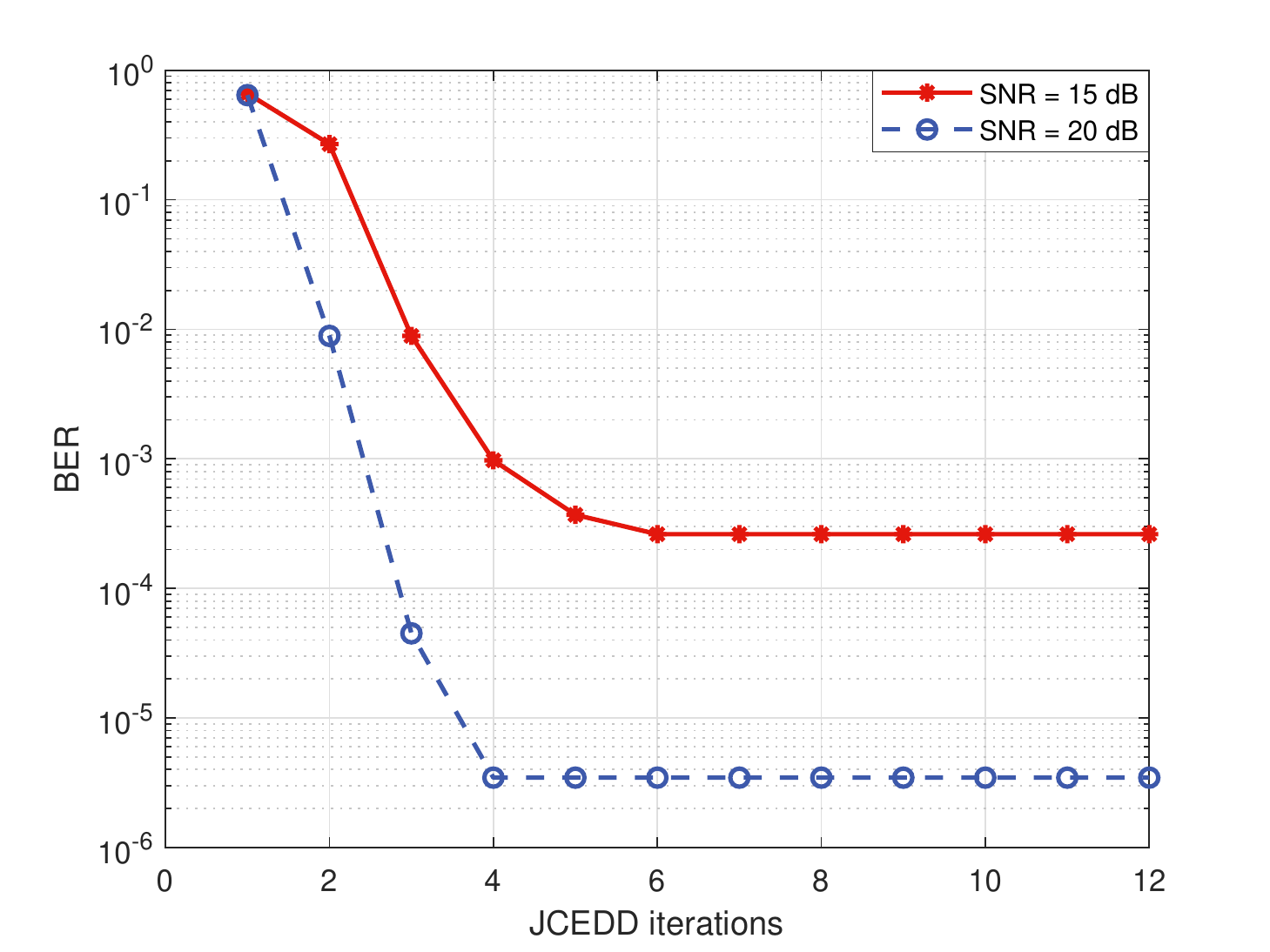}
  \end{minipage}
  }
  \caption{Performance of the proposed JCEDD scheme versus iterations: (a) NMSEs of the channel parameters;
  (b) BER. }
  \label{convergence_of_JCEDD_with_QPSK}
\end{figure}

Then we focus on the proposed JCEDD scheme.
In \figurename{ \ref{convergence_of_JCEDD_with_QPSK}}, the NMSEs and the BER versus the iteration index of JCEDD scheme are presented, where both $\text{SNR} = 15$ dB and $20$ dB are considered, and the power gap factor $\Delta = 6$ dB.
In addition, 4QAM modulation is adopted.
Under both the two SNRs, the NMSEs of the parameters and the BER all decrease with the iteration number increases.
After a few number of iterations, the curves for all the NMSEs and the BER approach their steady states, which verify the convergence and effectiveness of our
proposed JCEDD scheme.
In addition, it can be observed that the case with higher SNR has better convergence, estimation and detection accuracy than that of lower SNR.

\begin{figure}[htbp]
  \centering
  \subfigure[]
  {\label{QPSKMSE_fulldata_addguard_vs_SNR}
  \begin{minipage}{70mm}
  \centering
  \includegraphics[width=70mm]{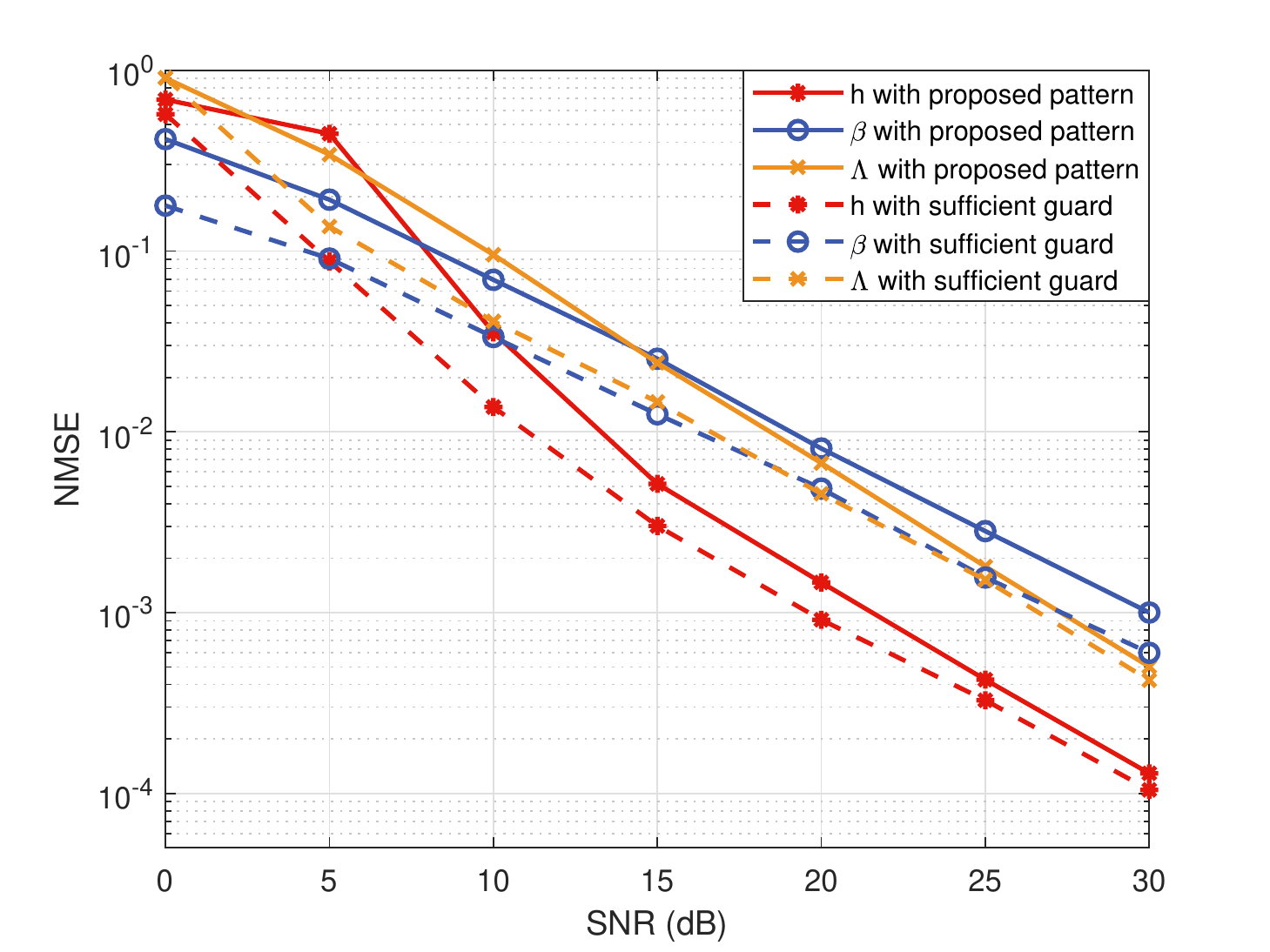}
  \end{minipage}
  }
  \subfigure[]
  {\label{QPSKBER_fulldata_addguard_vs_SNR}
  \begin{minipage}{70mm}
  \centering
  \includegraphics[width=70mm]{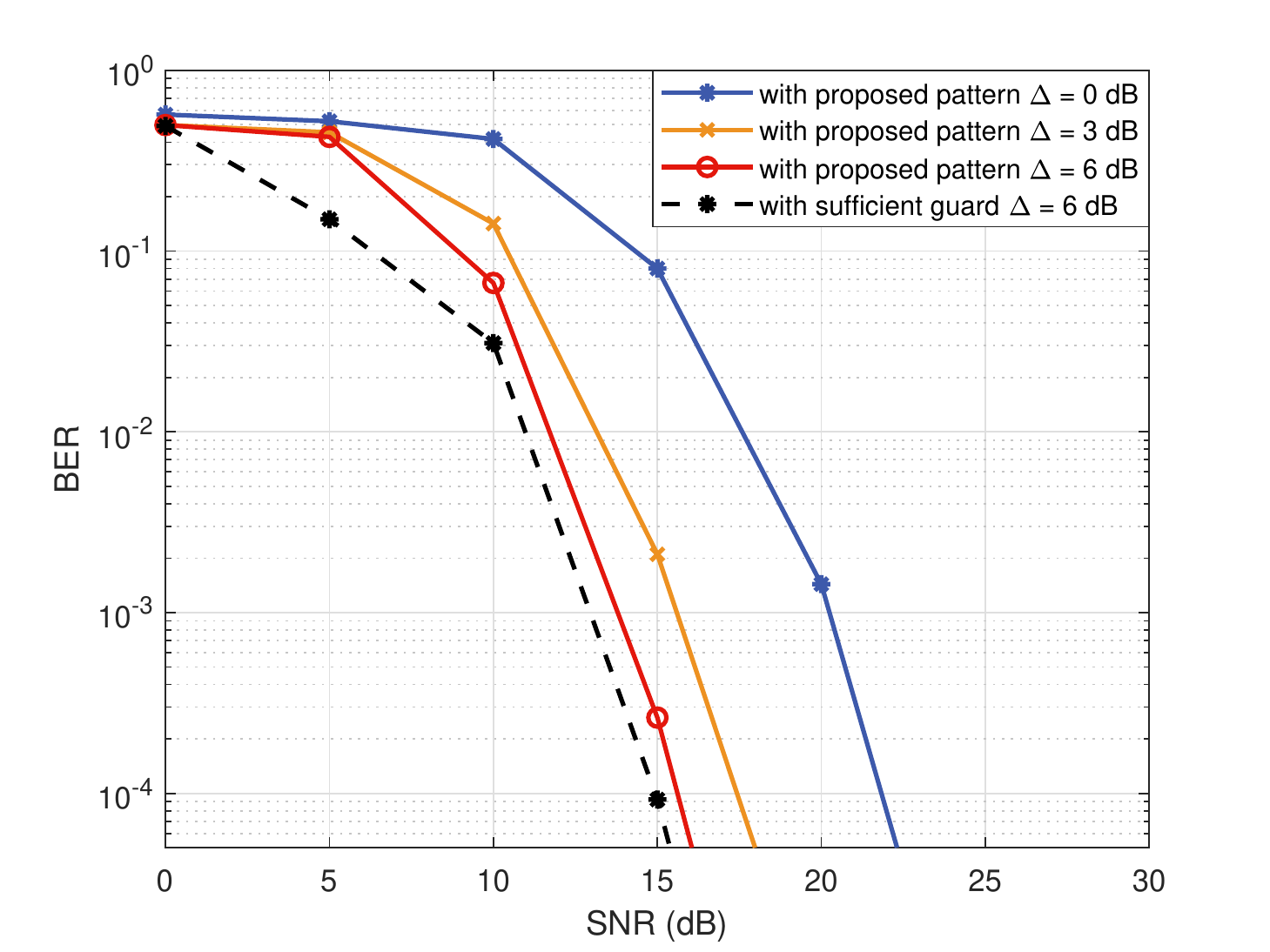}
  \end{minipage}
  }
  \caption{Performance of the proposed JCEDD scheme versus SNR: (a) NMSEs of the channel parameters with $\Delta = 6$ dB;
  (b) BER with different levels of power gap factors.}
  \label{performance_of_JCEDD_with_QPSK}
\end{figure}

\figurename{ \ref{performance_of_JCEDD_with_QPSK}} illustrates the NMSE of the channel parameters and BER performance of the JCEDD scheme versus SNR with 4QAM modulation.
Note that the case with sufficient guard grids is also taken into consideration as the baseline performance, where the signal pattern is illustrated in \figurename{ \ref{old_pattern2}}.
As shown in \figurename{ \ref{QPSKMSE_fulldata_addguard_vs_SNR}} and \figurename{ \ref{QPSKBER_fulldata_addguard_vs_SNR}}, the NMSEs of the channel parameters and BER exhibit worse performance when the SNR is lower than $5$ dB.
However, as SNR increases, the NMSEs decrease substantially.
Besides, the BER with $\Delta = 6$ dB reaches $10^{-4}$ around $\text{SNR} = 16$ dB.
Moreover, the NMSEs and the BER with the proposed pattern even approach those in the situation with sufficient guard when $\text{SNR} > 15$ dB.
This can be explained as follows.
When $\text{SNR}$ is sufficiently high, the data symbols are almost correctly detected, hence the data symbols in the interference area can be regarded as pilot for later iterations of the JCEDD scheme.
Since the number of acknowledged symbols becomes larger, the channel estimation accuracy and the BER performance can be improved accordingly.
In addition, we depict the BER with different levels of power gap factors in \figurename{ \ref{QPSKBER_fulldata_addguard_vs_SNR}}.
It is obvious that larger power gap can help improve the BER performance of data detection.
Besides, when $\Delta = 6$ dB, it can be checked that the BER with sufficient guard is always lower than that with the proposed pattern for low SNR.
However, even the pilot overhead is very low, the BER with our proposed pattern becomes very close to that with sufficient guard for high SNR, which shows the effectiveness of the proposed JCEDD scheme.

\begin{figure}[htbp]
  \centering
  \subfigure[]
  {\label{QPSKMSE_fulldata_addguard_CEBDD_JCEDD_vs_SNR}
  \begin{minipage}{70mm}
  \centering
  \includegraphics[width=70mm]{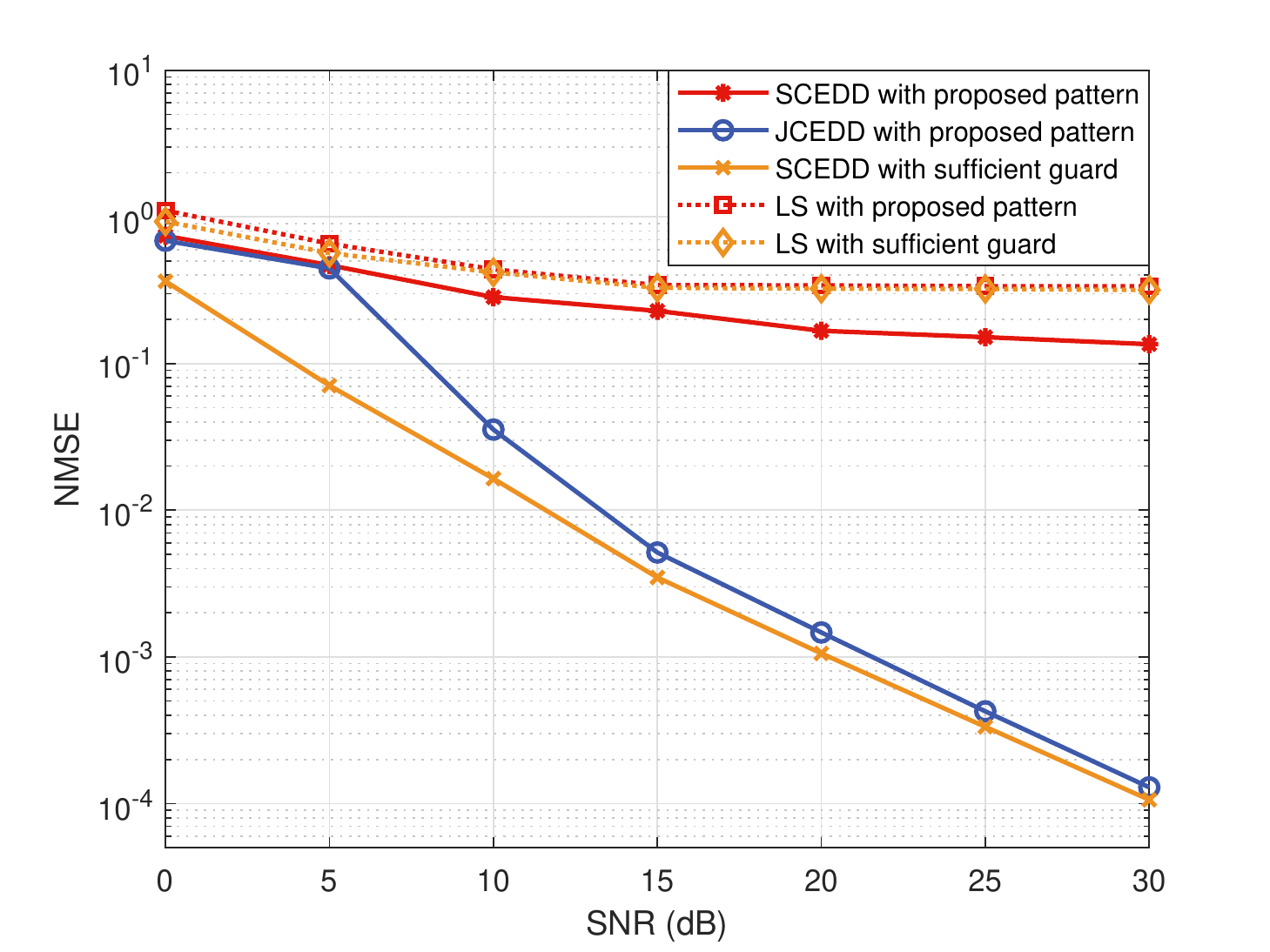}
  \end{minipage}
  }
  \subfigure[]
  {\label{QPSKBER_fulldata_addguard_CEBDD_JCEDD_vs_SNR}
  \begin{minipage}{70mm}
  \centering
  \includegraphics[width=70mm]{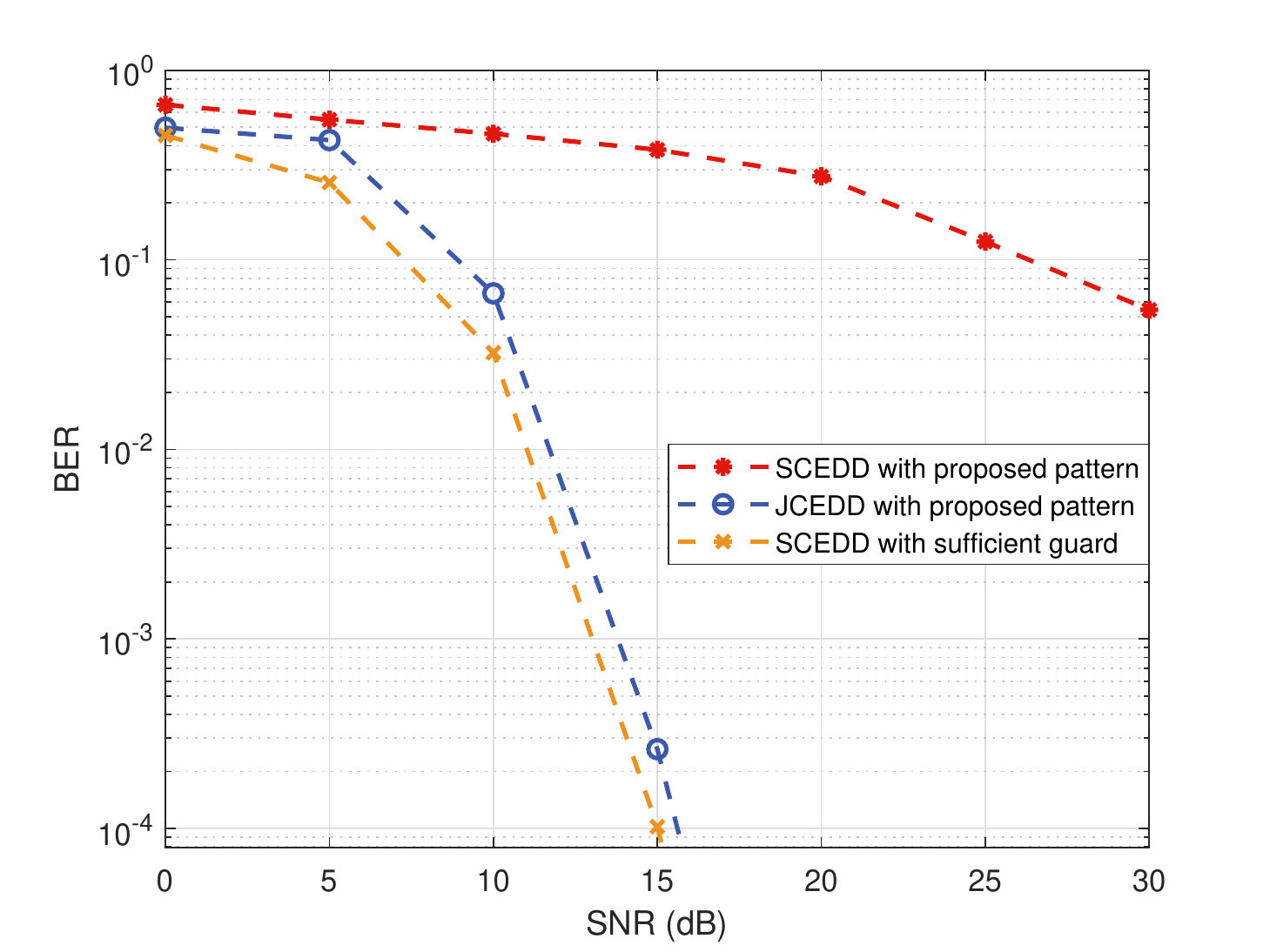}
  \end{minipage}
  }
  \caption{Performance of different situations and schemes versus SNR: (a) NMSEs of the equivalent channel gain;
  (b) BER.}
  \label{performance_of_different_situation_with_QPSK}
\end{figure}

Moreover, the performance improvement brought by the joint consideration is examined in \figurename{ \ref{performance_of_different_situation_with_QPSK}}, where 4QAM modulation and the power gap factor $\Delta = 6$ dB are implemented for all the cases.
Note that ``SCEDD'' represents ``separately channel estimation and data detection''.
Moreover, the equivalent channel gain estimation performance of least square (LS) estimator with the two symbol patterns in \figurename{ \ref{old_pattern}} is also illustrated.
From \figurename{ \ref{QPSKMSE_fulldata_addguard_CEBDD_JCEDD_vs_SNR}}, it can be observed that the LS channel estimation performance under the two symbol pattern are both significantly worse than that of the proposed scheme.
This is due to the interference between neighbored pilot symbols.
Besides, the performance of LS with sufficient guard is a little better than that with the proposed pattern. This is resulted from the elimination of the interference from data symbols.
Except for LS,
it is obvious that the performance of SCEDD scheme with sufficient guard has the best performance, and that with the proposed pattern has the worst.
This is due to the fact that severe interference between the pilot and data is not handled properly if the channel estimation and data detection are separately processed for the proposed pattern.
With such mutual interference, the accuracy of channel estimation can not be guaranteed, which will further effect the data detection subsequently.
However, the proposed JCEDD scheme can alleviate such interference and guarantee the performance.
From both \figurename{ \ref{QPSKMSE_fulldata_addguard_CEBDD_JCEDD_vs_SNR}} and \figurename{ \ref{QPSKBER_fulldata_addguard_CEBDD_JCEDD_vs_SNR}}, it can be observed that the performance of the proposed JCEDD scheme with the proposed pattern is very close to that of SCEDD scheme with sufficient guard, which exhibits the effectiveness and high spectral efficiency of the proposed JCEDD scheme.

\begin{figure}[htbp]
  \centering
  \subfigure[]
  {\label{many_modulationsMSE_240kmh_fullyplaced_vs_SNR}
  \begin{minipage}{70mm}
  \centering
  \includegraphics[width=70mm]{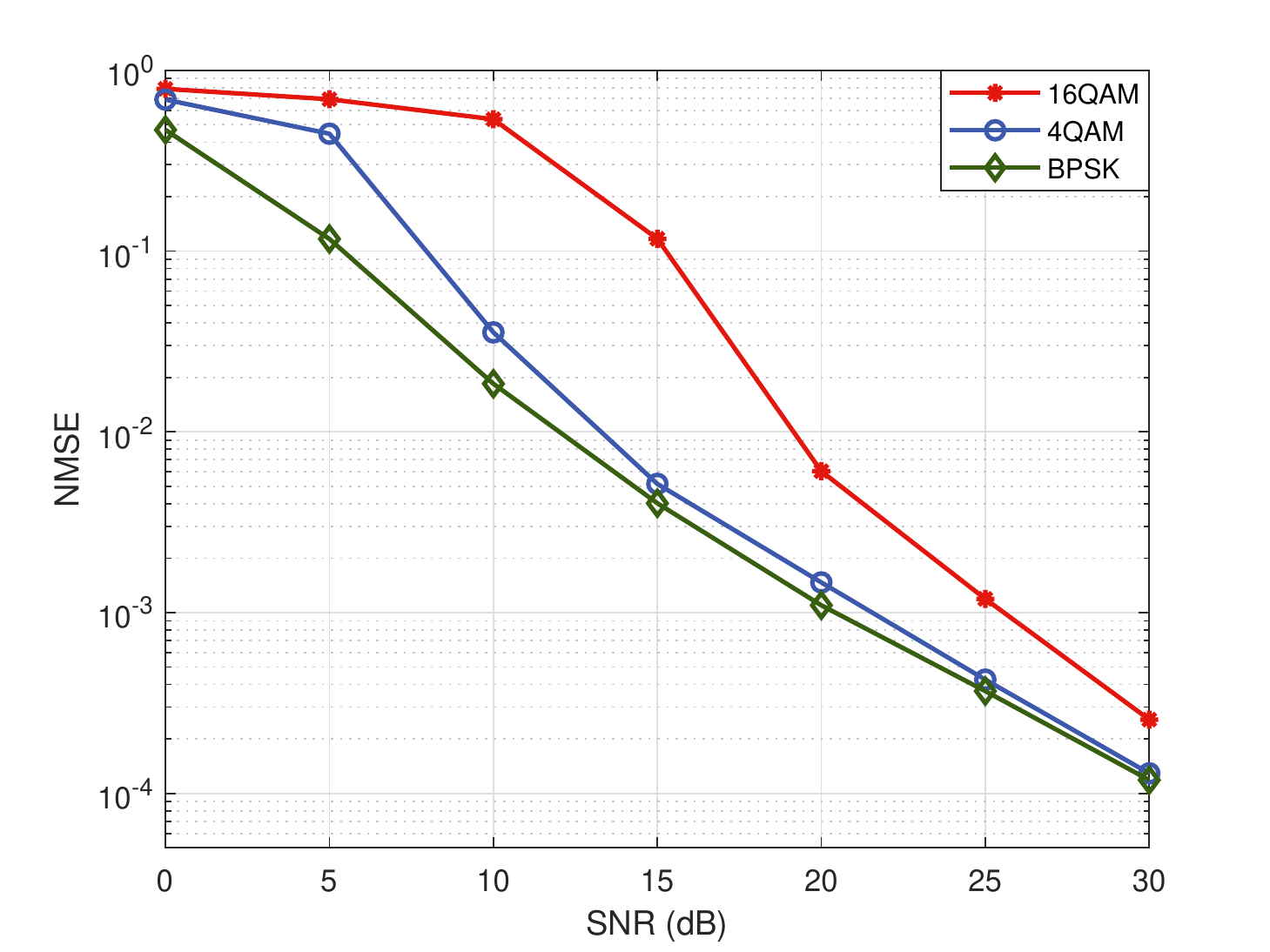}
  \end{minipage}
  }
  \subfigure[]
  {\label{many_modulationsBER_240kmh_fullyplaced_vs_SNR}
  \begin{minipage}{70mm}
  \centering
  \includegraphics[width=70mm]{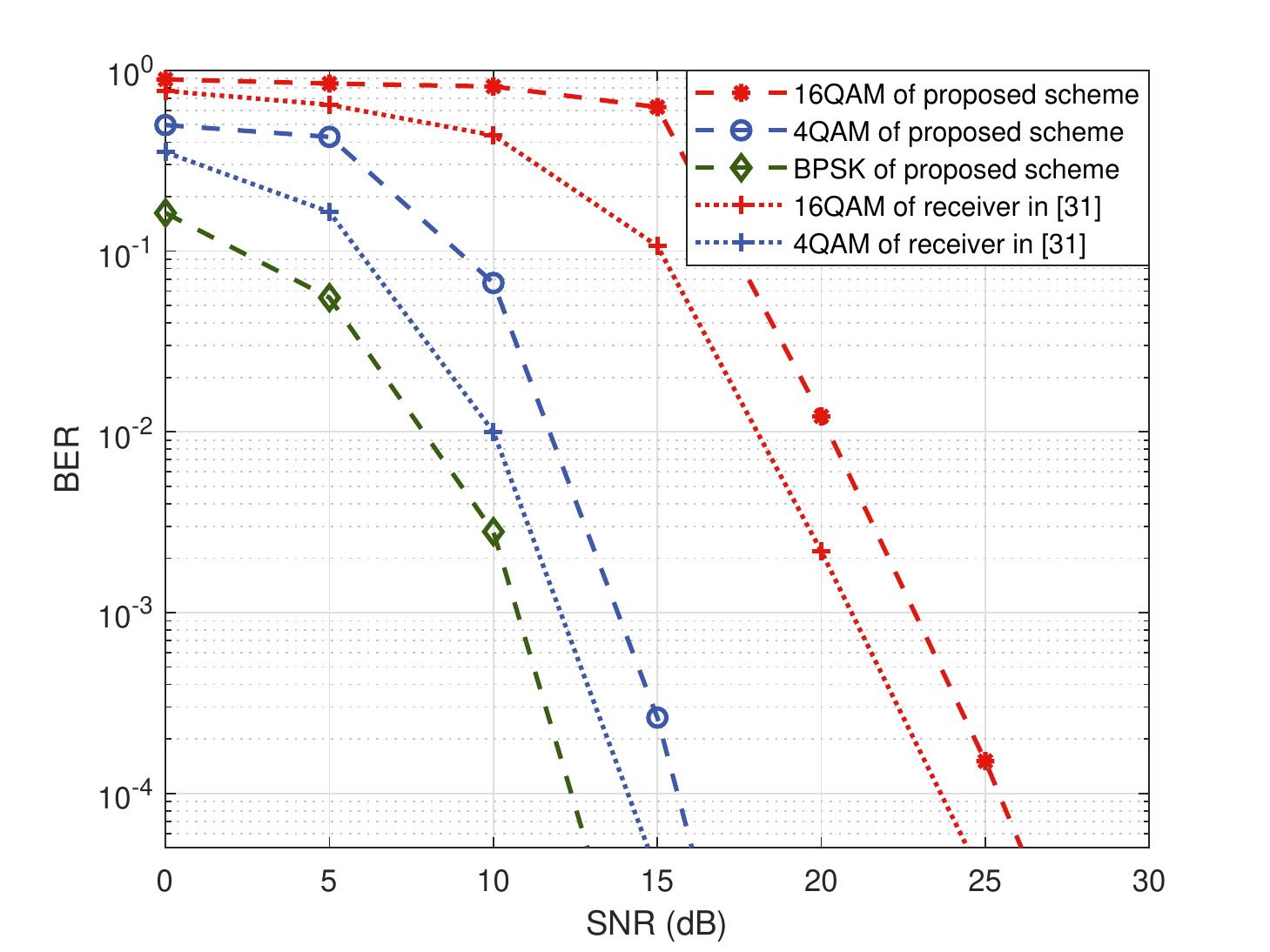}
  \end{minipage}
  }
  \caption{Performance of the proposed JCEDD scheme versus SNR with different modulation orders: (a) NMSE of the equivalent channel gain;
  (b) BER. }
  \label{many_modulations}
\end{figure}

In \figurename{ \ref{many_modulations}}, we compare the performance of the proposed JCEDD scheme with different modulation orders, where the power gap factor $\Delta = 6$ dB.
It can be checked in \figurename{ \ref{many_modulationsBER_240kmh_fullyplaced_vs_SNR}} that the BER increases gradually as the modulation order becomes higher.
Take $\text{SNR} = 15$ dB as an example, the BER with 4QAM modulation is slightly higher than $10^{-4}$, while that of 16QAM modulation is still very high.
This is due to the inaccuracy of the estimated CSI for 16QAM modulation shown in \figurename{ \ref{many_modulationsMSE_240kmh_fullyplaced_vs_SNR}}.
However, when the SNR goes higher, the NMSE of the equivalent channel gain with 16QAM rapidly approaches to that of 4QAM and BPSK.
With more accurate CSI, the BER of 16QAM modulation also decreases substantially,, and reaches $10^{-4}$ around $\text{SNR} = 25$ dB.
Besides, we further compared the BER of the proposed JCEDD scheme with the designed receiver in \cite{OTFS2}.
Note that the BER of the receiver in \cite{OTFS2} is based on the knowledge of precise CSI.
It can be observed that the BER of our proposed JCEDD scheme is gradually approaching that of the receiver with precise CSI in \cite{OTFS2} as the SNR increases.
This is due to the fact that the channel estimation is getting more accurate in our proposed scheme and can achieve similar BER performance of precise CSI receiver when the SNR is high.

\begin{figure}[htbp]
  \centering
  \subfigure[]
  {\label{QPSKMSE_fulldata_240_120kmh_480kmh_vs_SNR}
  \begin{minipage}{70mm}
  \centering
  \includegraphics[width=70mm]{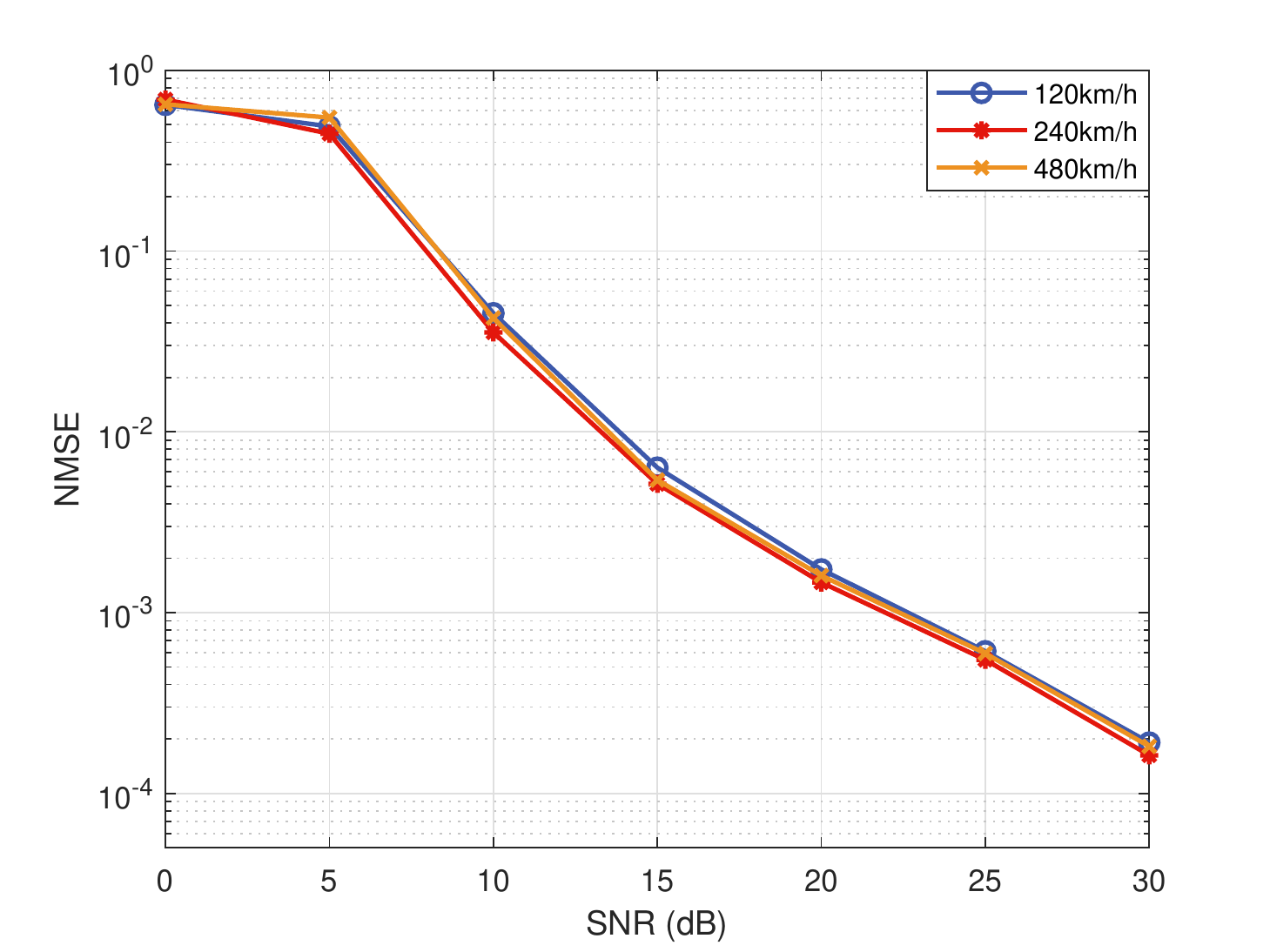}
  \end{minipage}
  }
  \subfigure[]
  {\label{QPSKBER_fulldata_240_120kmh_480kmh_vs_SNR}
  \begin{minipage}{70mm}
  \centering
  \includegraphics[width=70mm]{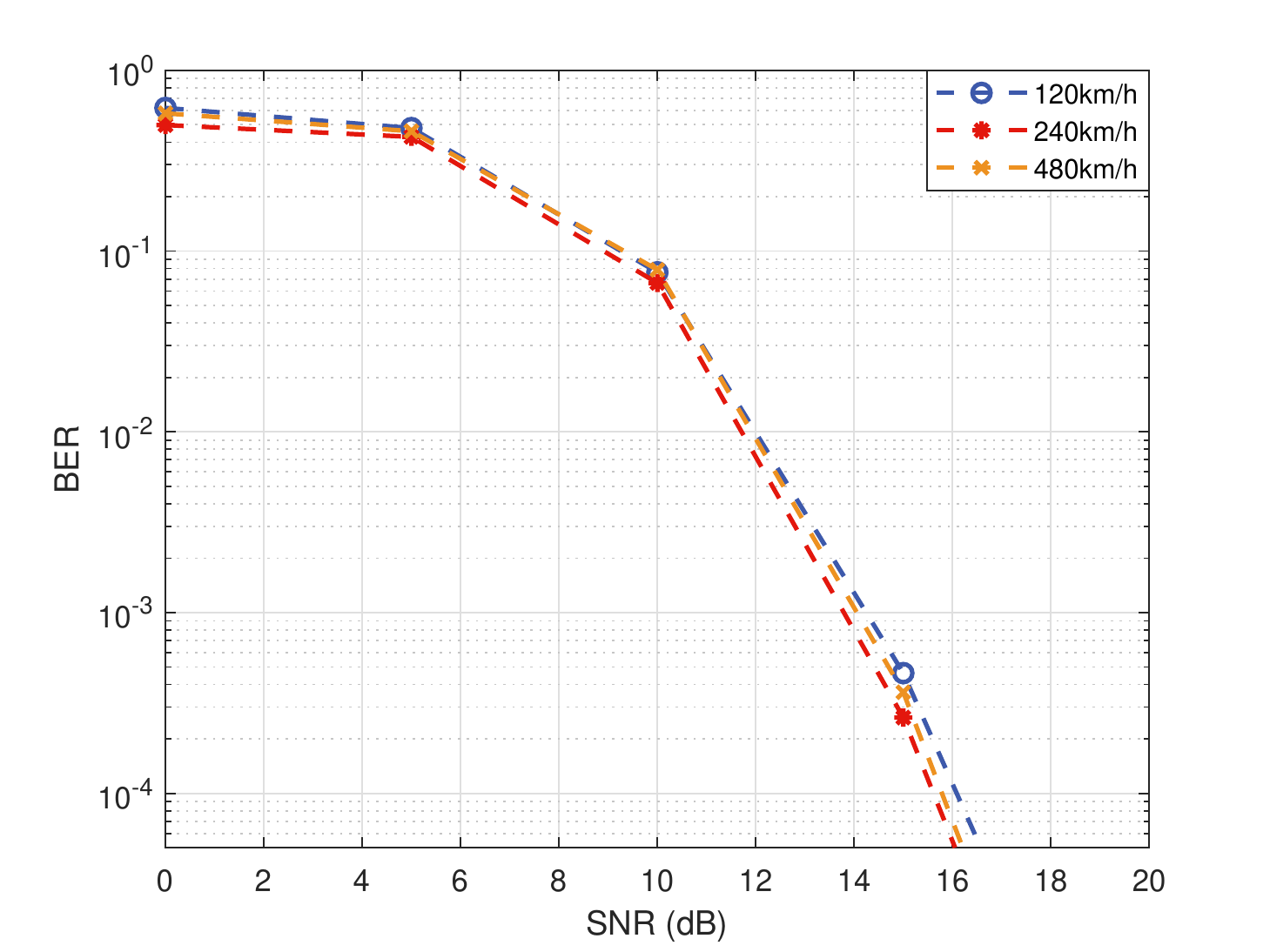}
  \end{minipage}
  }
  \caption{Performance of the proposed JCEDD scheme versus SNR with different user velocities: (a) NMSE of the equivalent channel gain; (b) BER.}
  \label{different_velocities}
\end{figure}

\figurename{ \ref{different_velocities}} shows the performance of the JCEDD scheme with different user velocities, where 4QAM modulation and $\Delta = 6$ dB are adopted.
It can be easily observed that both the estimation and detection performance are very similar for different velocities.
This is due to the reason that OTFS is very insensitive to the Doppler frequency shifts, where similar scattering paths can be distinguished even with different velocities.
Hence, similar estimation and detection results are finally observed for different user velocities.


\begin{figure}[htbp]
  \centering
  \subfigure[]
  {\label{new_patterns2}
  \begin{minipage}{70mm}
  \centering
  \includegraphics[width=70mm]{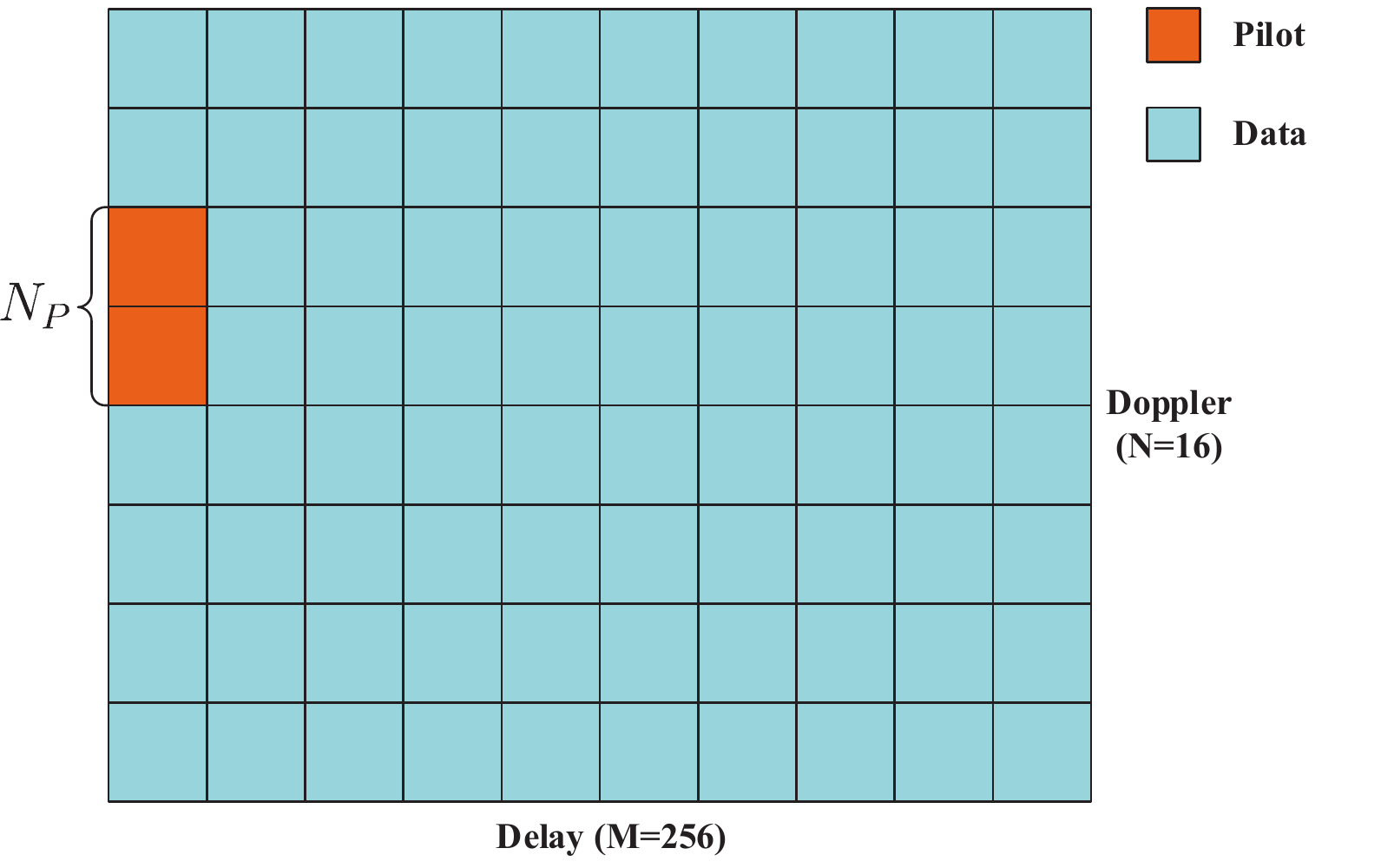}
  \end{minipage}
  }
  \subfigure[]
  {\label{new_patterns1}
  \begin{minipage}{70mm}
  \centering
  \includegraphics[width=70mm]{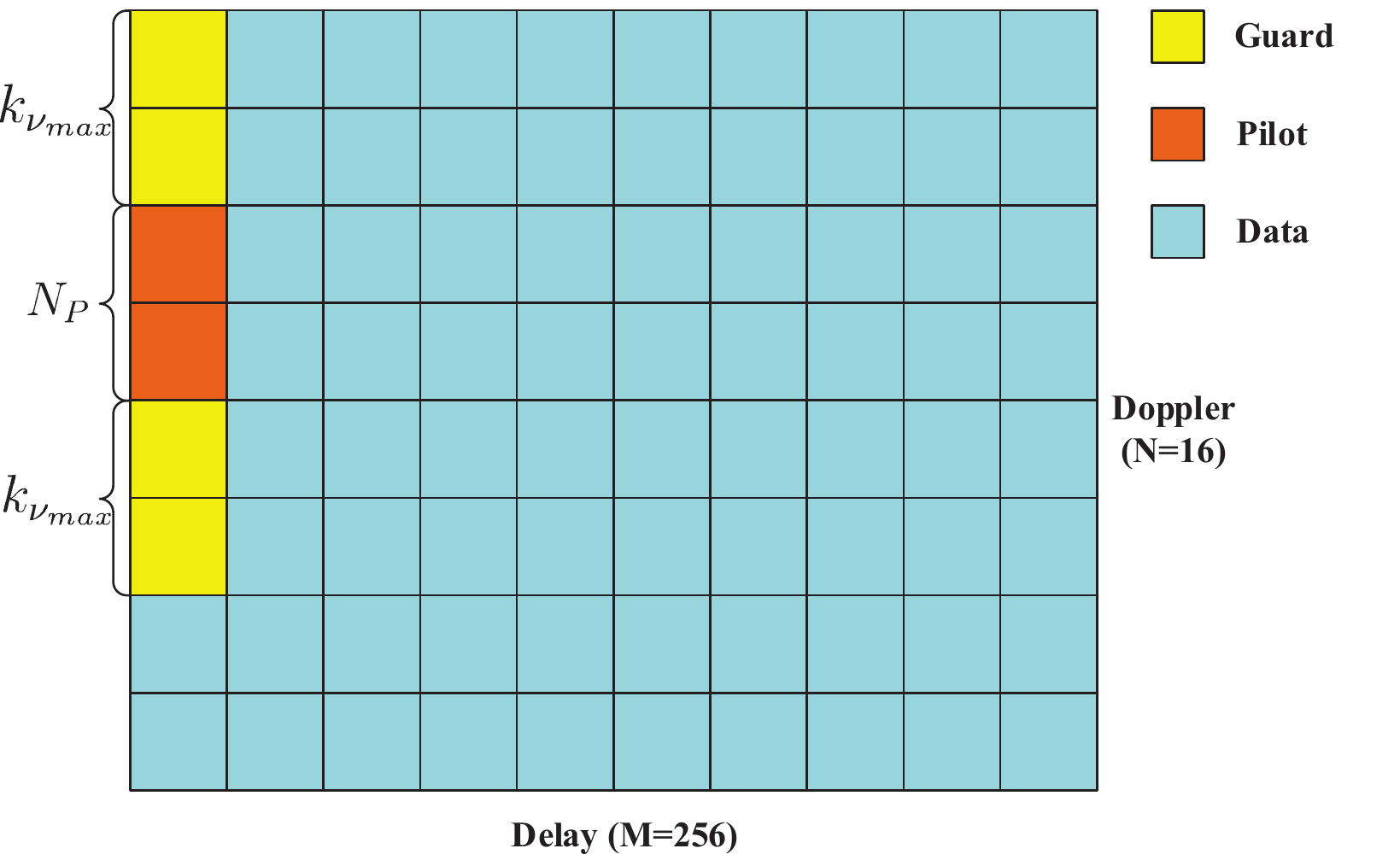}
  \end{minipage}
  }
  \caption{Patterns with only one column of pilot: (a) without guard; (b) reduced guard on Doppler domain.}
  \label{pattern_with_reduced_overhead}
\end{figure}

\begin{figure}[htbp]
  \centering
  \subfigure[]
  {\label{QPSKMSE_480kmh_truenoguard_ng10_20_guard2Dop_ng10_vs_SNR}
  \begin{minipage}{70mm}
  \centering
  \includegraphics[width=70mm]{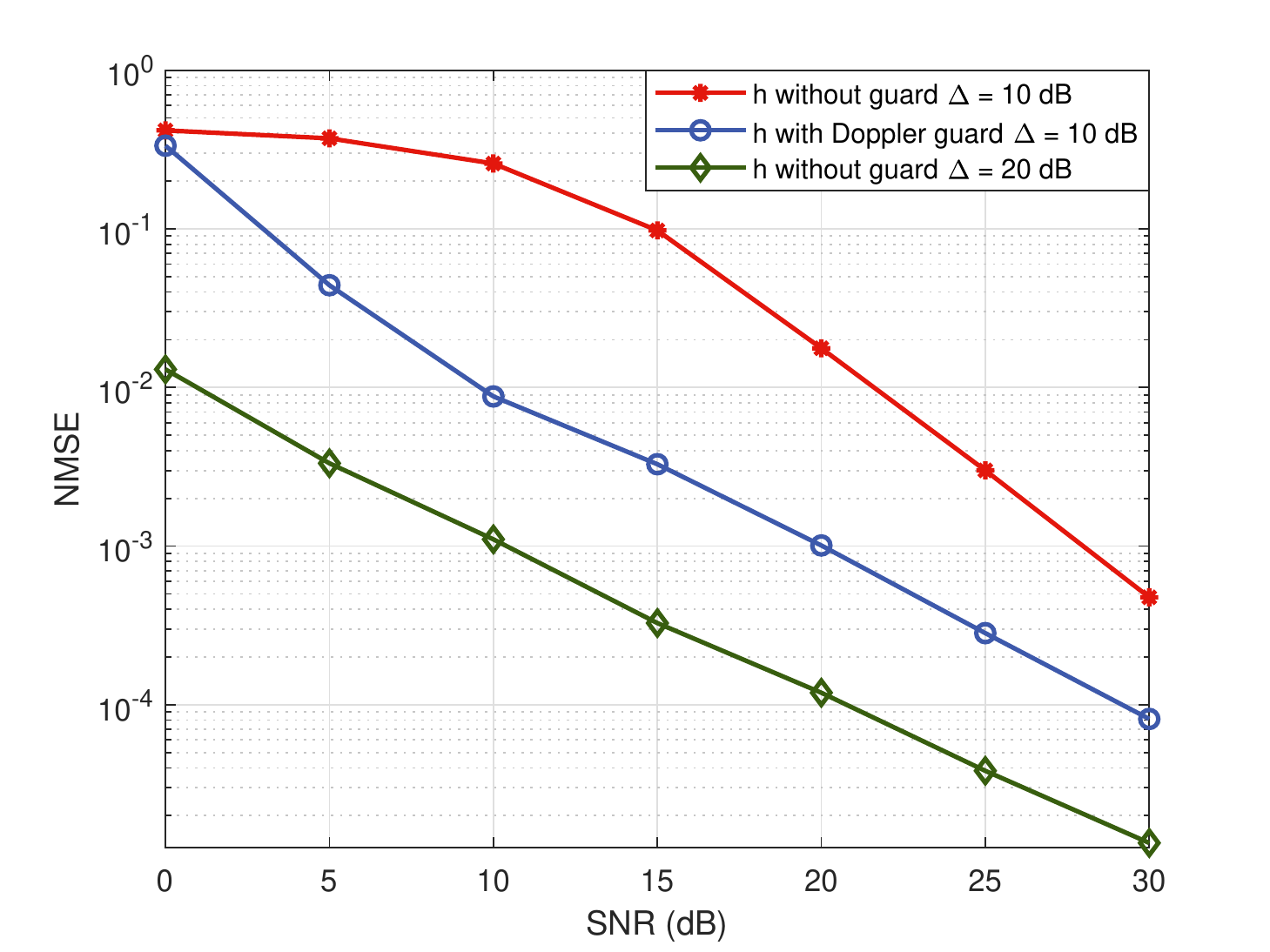}
  \end{minipage}
  }
  \subfigure[]
  {\label{QPSKBER_fulldata_240_120kmh_480kmh_vs_SNR}
  \begin{minipage}{70mm}
  \centering
  \includegraphics[width=70mm]{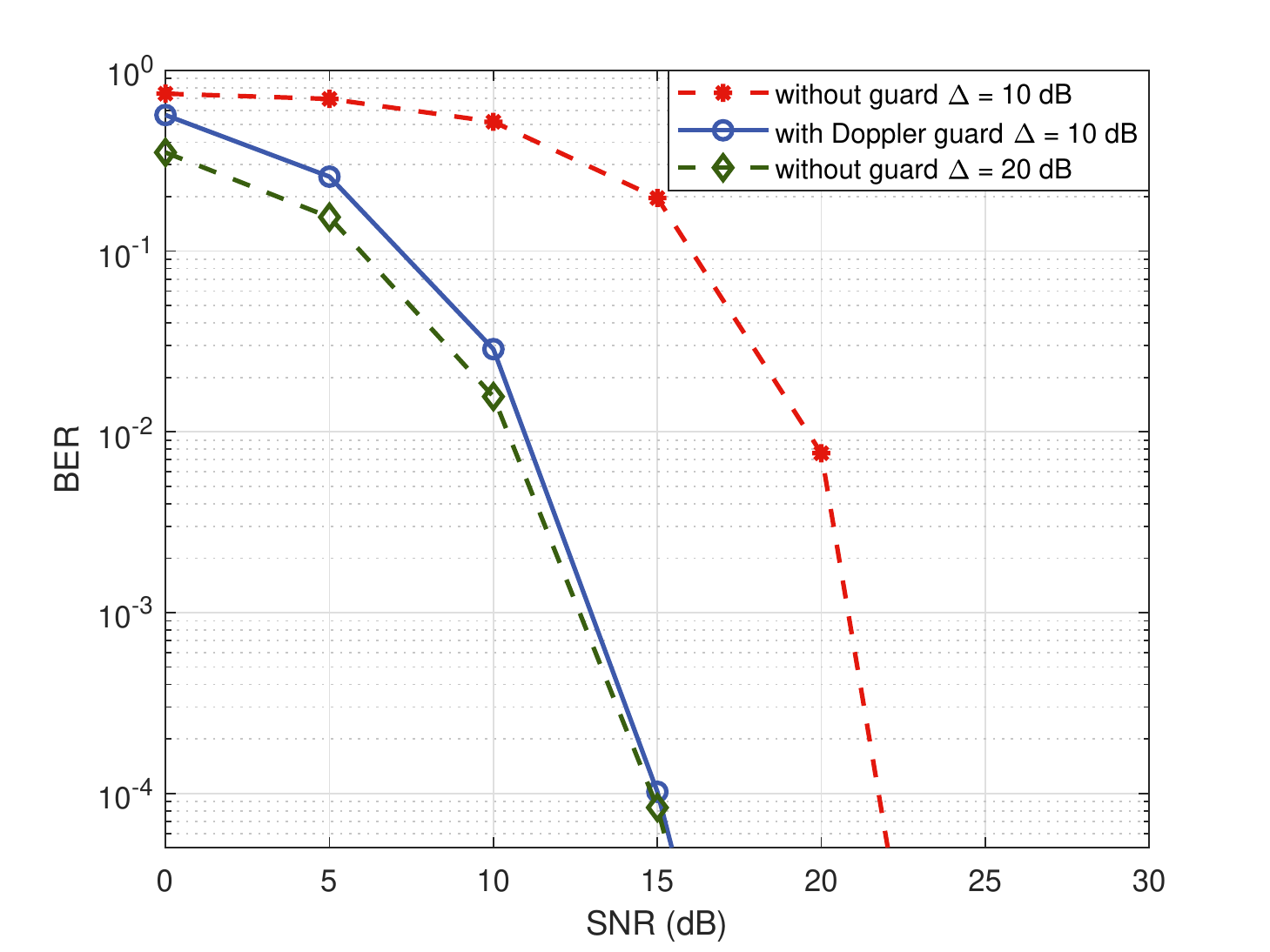}
  \end{minipage}
  }
  \caption{Performance versus SNR by using patterns with reduced pilot/guard overhead: (a) NMSE of the equivalent channel gain; (b) BER.}
  \label{reduced_overhead}
\end{figure}

Lastly, we examine the proposed JCEDD under extremely poor situations, where 4QAM modulation is adopted.
The signal patterns adopted here are shown in \figurename{ \ref{pattern_with_reduced_overhead}}, where we only place $N_P \times M_P = 4 \times 1$ pilot symbols.
Note that no guard grid is placed in \figurename{ \ref{new_patterns2}}, and only a few guard grids are placed in \figurename{ \ref{new_patterns1}} over the Doppler domain.
The performance of the JCEDD scheme is presented in \figurename{ \ref{reduced_overhead}}, where 4QAM modulation is adopted and the user velocity is $480$ km/h.
It can be observed that the performance with the pattern in \figurename{ \ref{new_patterns2}} is very poor when $\Delta = 10$ dB due to the severe mutual interference between the pilot and data.
Here, we propose two ways to improve the performance.
One is to rise the power gap factor to $20$ dB, and the other is to adopt the pattern in \figurename{ \ref{new_patterns1}}.
It can be verified that both the two ways can significantly improve the performance of our JCEDD scheme.

\begin{remark}
From the simulation results, it can be concluded that the performance of the proposed JCEDD scheme depends on both the number of guard grids and the power gap between pilot and data symbols.
The higher the power gap factor and the more guard grids, the better performance of the scheme.
However, it should be noticed that higher power gap can result in worse peak to average power ratio of the transmission system.
On the other hand, if sufficient pilot and guard grid are adopted, the pilot overhead will be very high, which is detrimental to the spectral efficiency.
Hence, a tradeoff should be reached between the estimation and detection performance, the power gap, and the pilot overhead.
\end{remark}

\subsection{Overhead Comparison between OTFS and OFDM}

\begin{figure}[htbp]
	\centering
	\includegraphics[width=90mm]{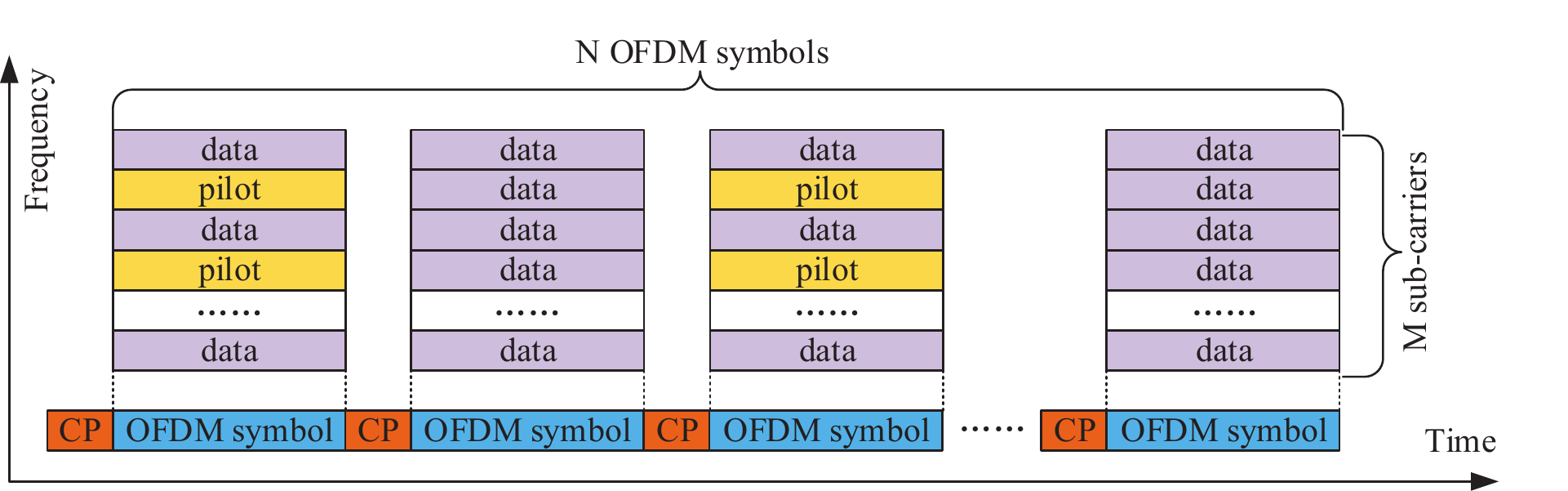}
	\caption{Time-frequency transmission signal pattern of OFDM.}
	\label{OFDM_pattern}
\end{figure}

To further verify the high spectrum efficiency of our proposed JCEDD scheme over high mobility scenario, we give overall overhead comparison between OTFS and OFDM, including CP, pilot symbols, and guards.
Since CP can only be analyzed over time domain, we will analyze and compare the overall overhead of OTFS and OFDM through their equivalent time cost.
Under any scale of user mobility, since we are to implement parameter extraction in time domain at the HRIS, it is obvious that the required preamble for HRIS beam design with both OFDM and OTFS are the same.
Besides, the interval of beam calibration is also equal for OFDM and OTFS.
Hence, the pilot overhead caused by the HRIS can be omitted in the comparison of the pilot overhead for OFDM and OTFS.
For notational simplicity, we define the maximum delay and Doppler frequency shift of the multi-path channel as $\tau_{max} = N_{\tau_{max}}T_s$ and $\nu_{max}$, respectively, where $T_s$ is the system sampling interval.
Then, the maximum dispersion grids over delay and Doppler domain will be $N_{\tau_{max}} = \tau_{max} / T_s$ and $k_{\nu_{max}}$, respectively.
We here analyze the overhead under the signal pattern in \figurename{ \ref{old_pattern1}} for OTFS and \figurename{ \ref{OFDM_pattern}} for OFDM within the duration of one OTFS block through their equivalent time cost.

For OTFS under the signal pattern in \figurename{ \ref{old_pattern1}}, it is set that $M_P = N_{\tau_{max}}$ and the length of pilot and guard over Doppler domain is equal to $N$.
Thus the overall equivalent time cost in the pattern of \figurename{ \ref{old_pattern1}} is $N_{\tau_{max}}N +N_{\tau_{max}}$, where the last item $N_{\tau_{max}}$ is the CP overhead of one OTFS block.
As for OFDM, assume that there are $P_M$ sub-carriers and $P_N$ OFDM symbols inserting pilot symbols.
Thus the pilot overhead is $P_M P_N$.
Besides, since CP should be added before each OFDM symbol, the CP overhead can be $N_{\tau_{max}}N$.
So the overall equivalent time cost of OFDM is $P_M P_N + N_{\tau_{max}}N$, which is higher than that of OTFS to some extent.
Moreover, we have introduced a signal pattern with extremely low overhead of pilot and guard in \figurename{ \ref{pattern_with_reduced_overhead}}.
It can be checked that the overall equivalent time cost of \figurename{ \ref{new_patterns2}} and \figurename{ \ref{new_patterns1}} is only $N_P+N_{\tau_{max}}$ and $2k_{\nu_{max}}+N_P+N_{\tau_{max}}$, respectively.
\figurename{ \ref{reduced_overhead}} illustrated that the proposed JCEDD scheme even with the pattern of \figurename{ \ref{new_patterns1}} also exhibits desired performance.

\begin{figure*}[!t]
\begin{align}\label{deriv_gamma1}
\frac{\partial {\gamma}_{k,l,p,q}^{(1)} }{\partial \beta_{\nu_p}}
&= \frac{\left(Q_0 Q_1(q) \jmath 2\pi (\frac{l-l_{\tau_p}}{MN}+1)e^{\jmath 2\pi (\frac{l-l_{\tau_p}}{MN} +1 ) \beta_{\nu_p}} - Q_0 \jmath 2\pi\frac{l-l_{\tau_p}}{MN} e^{\jmath 2\pi \frac{l-l_{\tau_p}}{MN} \beta_{\nu_p}}\right) (Q_2(q) e^{\jmath \frac{2\pi}{N} \beta_{\nu_p}} - 1)}
{(Q_2(q) e^{\jmath \frac{2\pi}{N} \beta_{\nu_p}} - 1)^2}
\notag \\
&- \frac{\left(Q_0 e^{\jmath 2\pi (\frac{l-l_{\tau_p}}{MN}) \beta_{\nu_p}} (Q_1(q) e^{\jmath 2\pi \beta_{\nu_p}} - 1) \right)Q_2(q) \jmath \frac{2\pi}{N} e^{\jmath \frac{2\pi}{N}\beta_{\nu_p} } }
{(Q_2(q) e^{\jmath \frac{2\pi}{N} \beta_{\nu_p}} - 1)^2},
\end{align}
\begin{align}\label{deriv_gamma1gamma1}
&\frac{\partial {\gamma}_{k,l,p,q}^{(1)}
({\gamma}_{k,l,p,q'}^1)^{*}}
{\partial \beta_{\nu_p}}
= \frac{-Q_1(q) \jmath 2\pi e^{\jmath 2\pi \beta_{\nu_p}} + Q_1^*(q')\jmath 2\pi e^{-\jmath 2\pi \beta_{\nu_p}}}
{1+ Q_2(q) Q_2^{*}(q')- Q_2(q) e^{\jmath \frac{2\pi}{N} \beta_{\nu_p}} - Q_2^*(q') e^{-\jmath \frac{2\pi}{N} \beta_{\nu_p}}}
\notag \\
&\ \ \ \ -
\frac{\left(-Q_2(q) \jmath \frac{2\pi}{N} e^{\jmath \frac{2\pi}{N} \beta_{\nu_p}} + Q_2^*(q') \jmath \frac{2\pi}{N} e^{-\jmath \frac{2\pi}{N} \beta_{\nu_p}}\right)
\left(1+Q_1(q)Q_1^*(q') -Q_1(q) e^{\jmath 2 \pi \beta_{\nu_p}} - Q_1^*(q') e^{-\jmath 2\pi \beta_{\nu_p}}\right)}
{\left(1+ Q_2(q) Q_2^{*}(q')- Q_2(q) e^{\jmath \frac{2\pi}{N} \beta_{\nu_p}} - Q_2^*(q') e^{-\jmath \frac{2\pi}{N} \beta_{\nu_p}}\right)^2}.
\end{align}
\hrulefill
\end{figure*}

\section{Conclusion}
In this paper, a JCEDD scheme for HRIS aided mmWave OTFS system was proposed.
To fully exploit the advantage of HRIS,
we firstly proposed a transmission structure, where the OTFS blocks were accompanied by a number of pilot sequences.
Within the duration of pilot sequences, partial HRIS elements were alternatively activated and the impinging signal was totally absorbed.
On the other hand, the HRIS was in passive mode within the duration of each OTFS block.
Then the time domain channel model was studied.
In addition, the received signal model at both the HRIS and the BS was presented.
Since the CSI between the user and the HRIS can be acquired with the help of pilot sequences, we proposed an HRIS beamforming design strategy to enhance the received signal strength at the BS.
For OTFS transmission, a JCEDD scheme was proposed.
In this scheme, we resorted to probabilistic graphical model, and designed a MP algorithm to simultaneously recover the data symbols and estimate the equivalent channel gain.
Besides, EM algorithm was employed to obtain channel parameters, i.e., the channel sparsity, the channel covariance, and the Doppler frequency shift.
By iteratively processing between the MP and EM algorithm, the delay-Doppler domain channel and the transmitted data symbols can be simultaneously obtained.
Simulation results were provided to demonstrate the validity of our proposed JCEDD scheme and its robustness to the user velocity.

\section*{Appendix A\\ Calculation for the Derivative of $Q(\beta_{\nu_p}, \widehat{k}_{\nu_p}^{(l-1)})$ with Respect to $\beta_{\nu_p}$}

Firstly, we focus on the derivative of ${\gamma}_{k,l,p,q}$ and ${\gamma}_{k,l,p,q}
({\gamma}_{k,l,p,q'})^{*}$ with respect to $\beta_{\nu_p}$.
If $p \leq l < M$, we have \eqref{deriv_gamma1} on the top of this page,
where $Q_0 = e^{\jmath 2\pi \frac{l-l_{\tau_p}}{MN} \hat{k}_{\nu_p}^{(l)}}$,
$Q_1(q) = e^{-\jmath 2\pi (\frac{N}{2} - q)}$,
$Q_2(q) = e^{-\jmath \frac{2\pi}{N} (\frac{N}{2}-q)}$,
and $\frac{\partial {\gamma}_{k,l,p,q}^{(1)}
({\gamma}_{k,l,p,q'}^1)^{*}}
{\partial \beta_{\nu_p}}$ is defined as \eqref{deriv_gamma1gamma1} on the top of this page.

On the other hand, when $0 \leq l < p$, it can be checked that
\begin{align}
\frac{\partial {\gamma}_{k,l,p,q}^{(2)} }{\partial \beta_{\nu_p}} =&
\phi(k,q,k_{\hat{\nu}_p})
\frac{\partial {\gamma}_{k,l,p,q}^{(1)} }{\partial \beta_{\nu_p}}, \label{deriv_gamma2}
\\
\frac{\partial {\gamma}_{k,l,p,q}^{(2)}
({\gamma}_{k,l,p,q'}^{(2)})^{*}}
{\partial \beta_{\nu_p}}
=& 
e^{\jmath \frac{2\pi}{N}\left([k-\hat{k}_{\nu_p}+q']_N- [k-\hat{k}_{\nu_p}+q]_N \right) }
\notag \\
&\times
\frac{\partial {\gamma}_{k,l,p,q}^{(1)}
({\gamma}_{k,l,p,q'}^{(1)})^{*}}
{\partial \beta_{\nu_p}}. \label{deriv_gamma2gamma2}
\end{align}

With \eqref{deriv_gamma1}-\eqref{deriv_gamma2gamma2}, the derivative of $Q(\beta_{\nu_p}, \hat{k}_{\nu_p}^{(l-1)})$ with respect to $\beta_{\nu_p}$ can be represented as \eqref{deriv_beta} on the top of the next page.
\begin{figure*}[!t]
\begin{align}\label{deriv_beta}
&\frac{\partial Q(\beta_{\nu_p}, \widehat{k}_{\nu_p}^{(l-1)})}{\partial \beta_{\nu_p}}
=
\frac{2}{\sigma_n^2} \Re\left\{ \sum_{(k,l) = (-\frac{N}{2},0)}^{(\frac{N}{2}-1, M-1)}
\sum_{q=0}^{N-1}
(y_{k,l})^*
\mathbb E \left\{\widetilde{h}_p | \mathbf y, \widehat{\boldsymbol \Xi}^{(l-1)}  \right\}
x_{k,l,p,q}
\frac{\partial {\gamma}_{k,l,p,q}}{\partial \beta_{\nu_p}} \right\}
\notag \\
&\quad\quad\quad\quad\quad- \frac{1}{\sigma_n^2} \sum_{(k,l) = (-\frac{N}{2},0)}^{(\frac{N}{2}-1, M-1)}
\sum_{q=0}^{N-1}
\sum_{q'=0}^{N-1}
\mathbb E\left\{\widetilde{h}_p \widetilde{h}^{*}_p| \mathbf y, \widehat{\boldsymbol \Xi}^{(l-1)}  \right\}
x_{k,l,p,q}
x_{k,l,p,q'}^{*}
\frac{\partial {\gamma}_{k,l,p,q}
({\gamma}_{k,l,p,q'})^{*}}
{\partial \beta_{\nu_p}}.
\end{align}
\hrulefill
\end{figure*}

\balance

\end{document}